\documentstyle[epsf,psfig,colordvi]{elsart}
\def\Cs {Cs$^{137}$ }

\def\NIM{\em Nucl. Instrum. Methods}

\def\NPB{{\em Nucl. Phys.} B}

\def\be{\begin{equation}}
\def\ee{\end{equation}}
\def\bea{\begin{eqnarray}}
\def\eea{\end{eqnarray}}
\begin{document}


\newcommand{\bdm}{\begin{displaymath}}
\newcommand{\edm}{\end{displaymath}}
\newcommand{\beqa}{\begin{eqnarray*}}
\newcommand{\eeqa}{\end{eqnarray*}}

\begin{frontmatter}
\title{Studies of the Response of the Prototype CMS Hadron Calorimeter,
Including Magnetic Field Effects, to Pion,
Electron, and Muon Beams}
\author[1a]{V.V.~Abramov},
\author[2a]{B.S.~Acharya},
\author[2a]{N.~Akchurin},
\author[17a]{I.~Atanasov},
\author[18a]{G.~Baiatian},
\author[4a]{A.~Ball}, 
\author[5a]{S.~Banerjee},  
\author[2a]{S.~Banerjee},
\author[6a]{P.~de~Barbaro},
\author[7a]{V.~Barnes},
\author[8a]{G.~Bencze},
\author[6a]{A.~Bodek},
\author[9a]{M.~Booke},
\author[6a]{H.~Budd},
\author[9a]{L.~Cremaldi},
\author[10a]{P.~Cushman},
\author[2a]{S.R.~Dugad},
\author[17a]{L.~Dimitrov},
\author[6a]{A.~Dyshkant},
\author[11a]{J.~Elias},
\author[1a]{V.N.~Evdokimov}, 
\author[4a]{D.~Fong},
\author[11a]{J.~Freeman},
\author[17a]{V.~Genchev}, 
\author[1a]{P.I.~Goncharov},
\author[11a]{D.~Green}, 
\author[5a]{A.~Gurtu}, 
\author[12a]{V.~Hagopian},
\author[17a]{P.~Iaydjiev},
\author[1a]{Yu.~Korneev},
\author[1a]{A.~Krinitsyn},
\author[6a]{G.~Krishnaswami}, 
\author[2a]{M.R.~Krishnaswamy},
\author[1a]{V.~Kryshkin},
\author[4a]{S.~Kunori}, 
\author[7a]{A.~Laasanen},
\author[12a]{D.~Lazic},
\author[19a]{L.~Levchuk},  
\author[16a]{L.~Litov}, 
\author[2a]{N.K.~Mondal},
\author[5a]{T.~Moulik}, 
\author[2a]{V.S.~Narasimham},
\author[19a]{A.~Nemashkalo}, 
\author[3a]{Y.~Onel},
\author[16a]{P.~Petrov},
\author[13a]{Yu.~Petukhov},
\author[17a]{S.~Piperov}
\author[19a]{V.~Popov}, 
\author[9a]{J.~Reidy},
\author[11a]{A.~Ronzhin},
\author[14a]{R.~Ruchti},
\author[15a]{J.B.~Singh},
\author[7a]{Q.~Shen},
\author[18a]{A.~Sirunyan},
\author[4a]{A.~Skuja},
\author[6a]{E.~Skup},
\author[19a]{P.~Sorokin}, 
\author[5a]{K.~Sudhakar}, 
\author[9a]{D.~Summers},
\author[8a]{F.~Szoncso},
\author[1a]{S.I.~Tereshenko}, 
\author[10a]{C.~Timmermans},
\author[5a]{S.C.~Tonwar},
\author[1a]{L.~Turchanovich}, 
\author[13a]{V.~Tyukov},
\author[13a]{A.~Volodko},
\author[13a]{A.~Yukaev},
\author[1a]{A.~Zaitchenko},
\author[19a]{A.~Zatserklyaniy}
\collab{(The CMS-HCAL Collaboration)}
\address[8a]{CERN, Geneva, Switzerland}
\address[13a]{Joint Institute for Nuclear Research, Dubna, Russia}
\address[11a]{Fermi National Accelerator Laboratory, Batavia, IL  60510}
\address[12a]{Florida State University, Tallahassee, FL 32306}
\address[3a]{University of Iowa, Iowa City, IA 52242}
\address[19a]{National Science Center Kharkov Institute of Physics and
Technology, Kharkov, Ukraine}
\address[4a]{University of Maryland, College Park, MD 20742}
\address[10a]{University of Minnesota, Minneapolis, MN 55455}
\address[9a]{University of Mississippi, University, MS 38677}
\address[14a]{University of Notre Dame, Notre Dame, IN 46556}
\address[15a]{Panjab University, Chandigarh, India}
\address[1a]{Institute for High Energy Physics, Protvino, Russia}
\address[7a]{Purdue University, West Lafayette, IN 47907}
\address[6a]{University of Rochester, Rochester, NY 14627}
\address[16a]{University of Sofia, Sofia, Bulgaria}
\address[17a]{Institute for Nuclear Research and Nuclear Energy, Sofia,
Bulgaria}
\address[18a]{Yerevan Physics Institute, Armenia}
\address[5a]{EHEP Group, Tata Institute of fundamental Research, Mumbai, India}
\address[2a]{HECR Group, Tata Institute of Fundamental Research, Mumbai, India}


\begin{abstract}

We report on the response of a prototype
CMS hadron calorimeter module to charged particle beams of pions, muons,
and electrons with momenta up to 375 GeV/c.
The data were taken at the H2 and H4 beamlines at CERN
in 1995 and 1996. 
The prototype
sampling calorimeter used copper
absorber plates and  scintillator tiles
with wavelength shifting fibers for readout.
The effects  of a magnetic field of
up to 3 Tesla on the response  of the calorimeter to muons, electrons, and
pions are presented, and
the effects of an upstream lead tungstate crystal  
electromagnetic calorimeter on the linearity and energy resolution of the
combined calorimetric system  to hadrons are evaluated. 
The results are compared with Monte Carlo simulations and are
used to optimize the
choice of total absorber depth, sampling frequency,
and longitudinal readout segmentation.

                and nuclear physics experimental methods and instrumentation
  29.40.Mc       Scintillation detectors
\vskip 0.1in
Six Keywords: Hadron Sampling Calorimeters, Magnetic Field Effects. 
\vskip 0.1in
UR-1593, CMS Note 00-003, {\today}\\
submitted to NIM A\\
Please send the proofs to: Pawel de Barbaro, U. of Rochester\\
c/o Fermilab MS318, P.O. Box 500,
Batavia, IL 60510, USA\\
tel: 630-840-5639, fax: 630-840-6485,
email: barbaro@fnal.gov 
\end{abstract}
\end{frontmatter}
%
{\section{ The CMS Scintillator Tile-Fiber Hadron Calorimeter }}

The hadron calorimeter (HCAL) for the Compact Muon Solenoid (CMS) Detector
will be used in the determination
of quark, gluon, and neutrino final state momenta by
measuring
the energies and directions of particle jets and the missing transverse
energy flow. The determination of the missing energy flow is
crucial in searches 
for new particles and phenomena, such
as possible supersymmetric partners of quarks and gluons.
Adequate granularity, resolution, and containment of particle showers
are essential in attaining
these performance goals and provide
one of the benchmarks in the design of the CMS hadron calorimeter.
In this communication we report
on test beam results~\cite{pawel}
used in the  optimization of the design of HCAL,
including choice of total absorber depth, sampling frequency,
and longitudinal readout segmentation.

The central CMS hadron calorimeter
~\cite{tp,tdr}
is located inside the 4~Tesla
coil of the CMS solenoid magnet (inner diameter of 5.9~m).
The central pseudorapidity range ( $|\eta|$ $< 3.0$ )
is covered by the barrel and endcap calorimeter systems. 
Segments of the crystal
PbWO$_4$  electromagnetic calorimeter with a silicon pre-shower detector
(in the endcap region only) are supported by the barrel and endcap hadron
 calorimeters.
The combined response of the
PbWO$_4$ crystal electromagnetic calorimeter and the hadron calorimeter
is used in the reconstruction of particle jets and missing
energy in the central pseudorapidity range.

The barrel and endcap
sampling hadron calorimeters are located inside the
4~Tesla field of the CMS solenoid.
Therefore, the calorimeters are constructed with
non-magnetic material. The absorber and structural elements
are made out of cartridge brass (70\% Cu/ 30\% Zn), and
stainless steel plates, respectively. 
Cartridge brass is easy to machine and its
hadronic interaction length is approximately 10\% shorter than iron.
The active sampling  elements are 3.7~mm
thick plastic scintillator tiles with wavelength-shifting (WLS) fiber
readout. The barrel
calorimeter inside the solenoid is relatively thin (about 7 interaction
lengths at $\eta = 0$).
To ensure adequate sampling depth
for $|\eta|$ $ < 1.4 $,  a hadron outer calorimeter is
installed outside the solenoid.
The outer calorimeter utilizes the solenoid magnet as an additional
absorber
equal to 1.4/sin($\theta$) interaction lengths and
is used to identify and quantify the contribution from
late starting showers.

\subsection{Design Requirements}

The design of the central hadron calorimeter requires good hermeticity, good
transverse granularity, moderate energy resolution, and sufficient depth for
hadron shower containment. We have chosen a lateral granularity of
$\Delta \eta \times \Delta \phi = 0.087 \times 0.087 $ for $|\eta| < 2.0 $.
This granularity is sufficient to insure good di-jet separation and mass
resolution.
The calorimeter readout 
is required to have a dynamic range from 20 MeV to 3 TeV. The
sensitivity at the low end allows for
 the observation of single muons in a calorimeter tower
for calibration and trigger redundancy.
The scale at the high end is set by the maximum
energy expected to be deposited by a jet in a single calorimeter tower.

Initial simulations of the CMS calorimeter indicate
that a  resolution of 
$\sigma_E/E = 120\%/\sqrt{E}$  ${\oplus}$  5\% 
for single incident hadrons is sufficient. In this case,
the energy  resolution for a jet of particles
between 50 GeV and 3 TeV is  not degraded by the measurement in the
calorimeter, when
 other fluctuations which are inherent in jets~\cite{lhcc_ref}
are also considered.

\subsection{Magnetic Field Effects}

Particular attention must be paid to the effects
of the high magnetic field on the response of the calorimeter. The
intrinsic shower
energy development and the containment of hadron showers have been
shown not to be affected by the high
magnetic field. However, the scintillator itself
exhibits an increased response in a magnetic field.
The increased response originates from two sources: (1)
an intrinsic brightening of the scintillator
($\approx$ 5\%) for B fields above 0.3~Tesla,  and (2) a geometric
effect due to the increased path lengths
of soft electrons in the scintillator. The latter effect
depends on the specific orientation of the magnetic field relative to
the calorimeter absorber plates.

\subsection{The Test Beam Program }

A test beam program was initiated in 1994~\cite{kryshkin} and was
continued in 1995 and 1996. A test module of
moveable brass absorber plates and scintillator tile-fiber sampling was
exposed to negative
 hadrons, electrons, and muons in the CERN test beam over a
large energy range. During this period the effects of the 
following sources on the response of the calorimeter
module were investigated:

\begin{itemize}

\item Magnetic field (scintillator ``brightening" and geometric effects)
\item Absorber thickness (optimization of resolution versus containment)
\item Absorber depth (energy containment)
\item Absorber non-uniformity (presence of the magnet coil)
\item Crystal electromagnetic calorimeter  (e/$\pi$ effects)
\end{itemize}

Data were taken both with and without a prototype lead
tungstate crystal electromagnetic calorimeter
(ECAL) placed upstream of the HCAL module.
These data have been used in the  optimization of the HCAL design,
including the choice of total absorber depth, sampling frequency,
and longitudinal readout segmentation. 

The test beam data show that the presence of 
the crystal electromagnetic calorimeter in front of HCAL
degrades the linearity and
resolution of the combined calorimeter system.
However,  a large fraction of this
degradation
can be corrected for by applying constant energy-independent
weighting factors  to the various longitudinal readout segments
of HCAL. The response of the combined calorimeter system using these
optimized weights meets the design requirements for
resolution and containment of hadron showers.

The experimental setup, various types
of studies, and data sets are described in section~2.
The measurement of the effects of a magnetic field
on the response of HCAL to muons, electrons, and pions is presented in
section~3.
The performance of HCAL and the
combined ECAL+HCAL system is discussed in sections~4 and~5.
Extraction of the intrinsic ratio of the response of HCAL to the
electromagnetic and hadronic components of the shower, $e/h$, is
presented in section~6.
The optimization studies leading to the choices
for the total absorber depth, absorber sampling frequency, and
longitudinal readout segmentation is presented in section~7.
A comparison of the test beam results with Monte Carlo
simulations is discussed in section~8. Section~9 gives 
a summary of the results and conclusions.

\section{Experimental Setup}

\begin{figure}
\begin{center}
\epsfxsize=4.5in
\mbox{\epsffile[45 45 576 729]{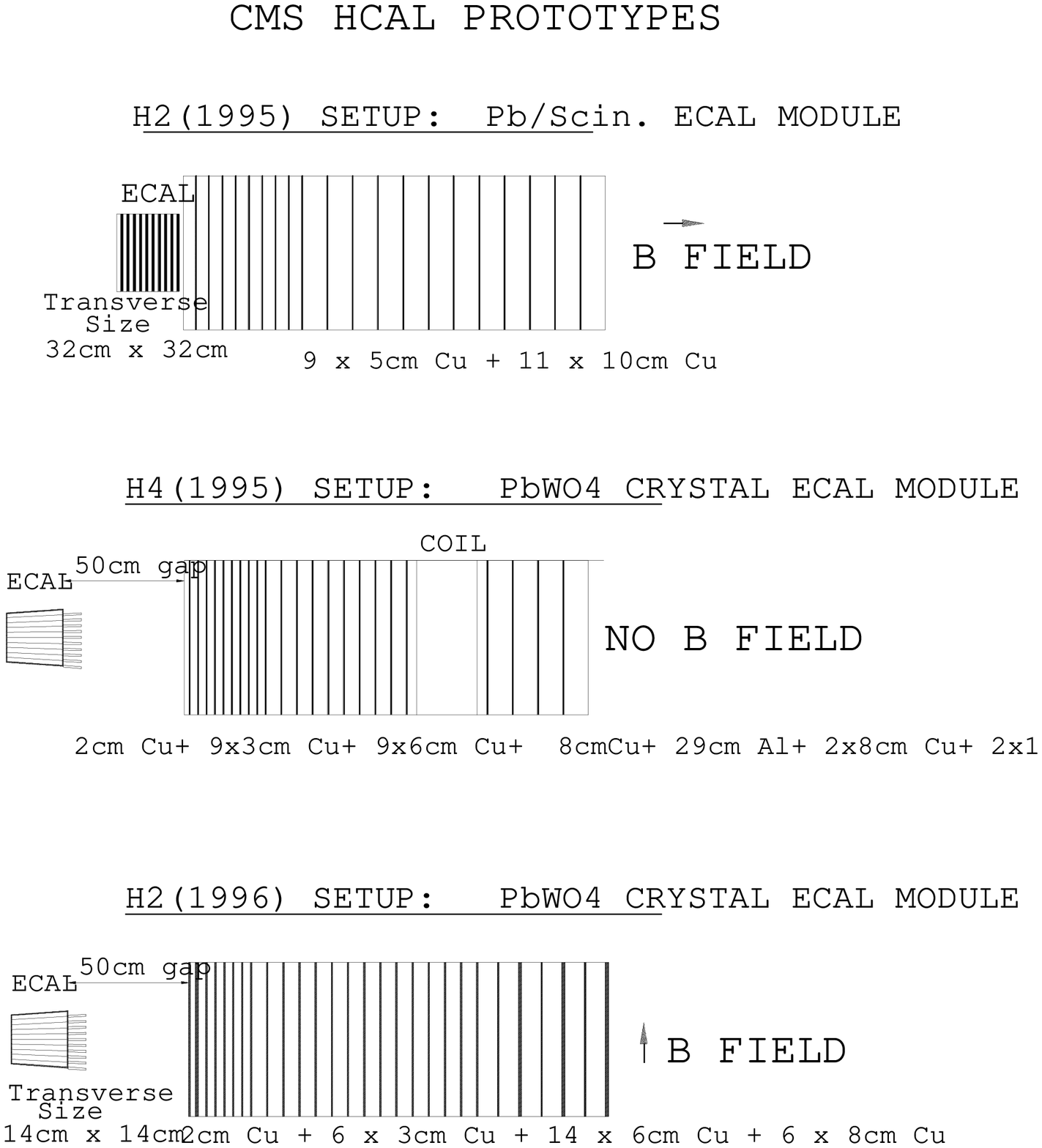}}
\caption{Schematic drawing of the  calorimeter modules used
during the 1995 and 1996 tests in the H2 and H4  beamlines.
A lead-scintillator sampling ECAL detector is used in the H2(1995) setup.
A 7$\times$7 matrix of
PbWO$_{4}$ crystal ECAL is used in the H4(1995) and
H2(1996) setups.
}
\label{nim98_2}
\end{center}
\end{figure}

The combined ECAL+HCAL calorimetric system for CMS has been tested in 1995
and 1996  at the H2 and H4 beamlines at CERN. Data were taken
using beams of muons,
electrons, and hadrons, ranging in momenta from 15 to 375 GeV/c.
These test beam prototypes of HCAL are based on the 
hanging file~\cite{kryshkin} structure,  in which copper alloy~\cite{alloy}
 absorber plates 
($\lambda_{INT}$(Cu)=15.06 cm)
varying in thickness from 2~cm to 10~cm are interspersed with
scintillator
tiles read out with WLS fibers~\cite{cdf_end_plug_upgrade_book}. 
Each scintillator tile is read out
independently with a photomultiplier tube (PMT)
located at the end of 10~m long optical cable. 
The total interaction depth 
of the HCAL prototypes corresponds to 10.3 $\lambda_{INT}$ (H2, 1995 setup),
8.5 $\lambda_{INT}$ (H4, 1995 setup),
and  10.1 $\lambda_{INT}$ (H2, 1996 setup).
The transverse size of the prototype scintillator tiles
is 64~cm$\times$64~cm.
Details of the H2(1995), H4(1995), and
H2(1996) experimental setups are shown in Figure~\ref{nim98_2}. 
The longitudinal segmentation  of each of the modules
is described in
Tables~\ref{h2_1995} through~\ref{h2_1996}.

In the H2(1995) run, the HCAL prototypes have been tested
with the detector placed inside a large 3 Tesla magnet~\cite{ref_magnet}.
The orientation of the magnetic field, with B field lines perpendicular
to the scintillator planes (magnetic field parallel to the beam) 
corresponds to the endcap configuration of a typical collider detector.
The H2(1995) tests include an ECAL module consisting
of a lead-scintillator
sampling calorimeter, with 10 layers of 1.6~cm Pb interspersed
with 6~mm thick scintillator plates. The transverse size
of the lead-scintillator sampling ECAL calorimeter
for the H2(1995) is 32~cm$\times$32~cm.

During the H4(1995) run,
the combined lead tungstate crystal ECAL+HCAL CMS prototype calorimeter
has been tested 
with no magnetic field.
The prototype H4(1995) ECAL detector
consists of a matrix of 7$\times$7 PbWO$_{4}$ crystals, 
each 23~cm long (25.8 X$_0$, 
1.1~$ \lambda_{INT}$), and 2~cm$\times$2~cm in transverse dimensions.
The total transverse size of the ECAL module matrix is
14~cm$\times$14~cm.
The crystal ECAL module is placed 
approximately 50~cm upstream of the front face of the HCAL prototype.
The PbWO$_{4}$ crystals are read out by avalanche
photodiodes~\cite{ecal_edr} (APD) with
a gain approximately equal to 50.

\begin{figure}
\begin{center}
\epsfxsize=4.5in
\mbox{\epsffile{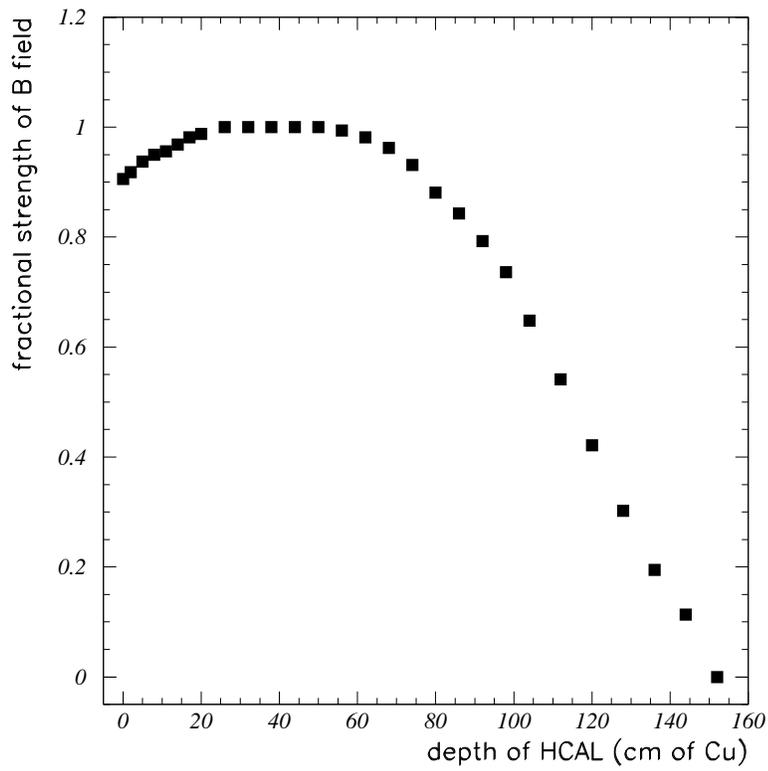}}
\caption{
Relative strength of the magnetic field, as a function of HCAL depth
(in cm of Cu) for the CMS H2(1996) test beam run.
}
\label{magnet}
\end{center}
\end{figure}

The combined crystal ECAL+HCAL calorimetric system 
has been tested again in 1996 at the CERN H2  beamline with a magnetic field. 
The 3 Tesla magnet is oriented in such a way that
the lines of the B field are parallel to the scintillator planes,
corresponding to the HCAL barrel configuration. 
For the first 5 interaction lengths of the calorimeter, 
the magnetic field is uniform to within 10\%. Figure~\ref{magnet} shows
the relative value of the magnetic field, as a function of depth in HCAL.

\subsection{The Relative  Calibration of  HCAL Layers}

\begin{figure}
\begin{center}
\epsfxsize=4.5in
\mbox{\epsffile{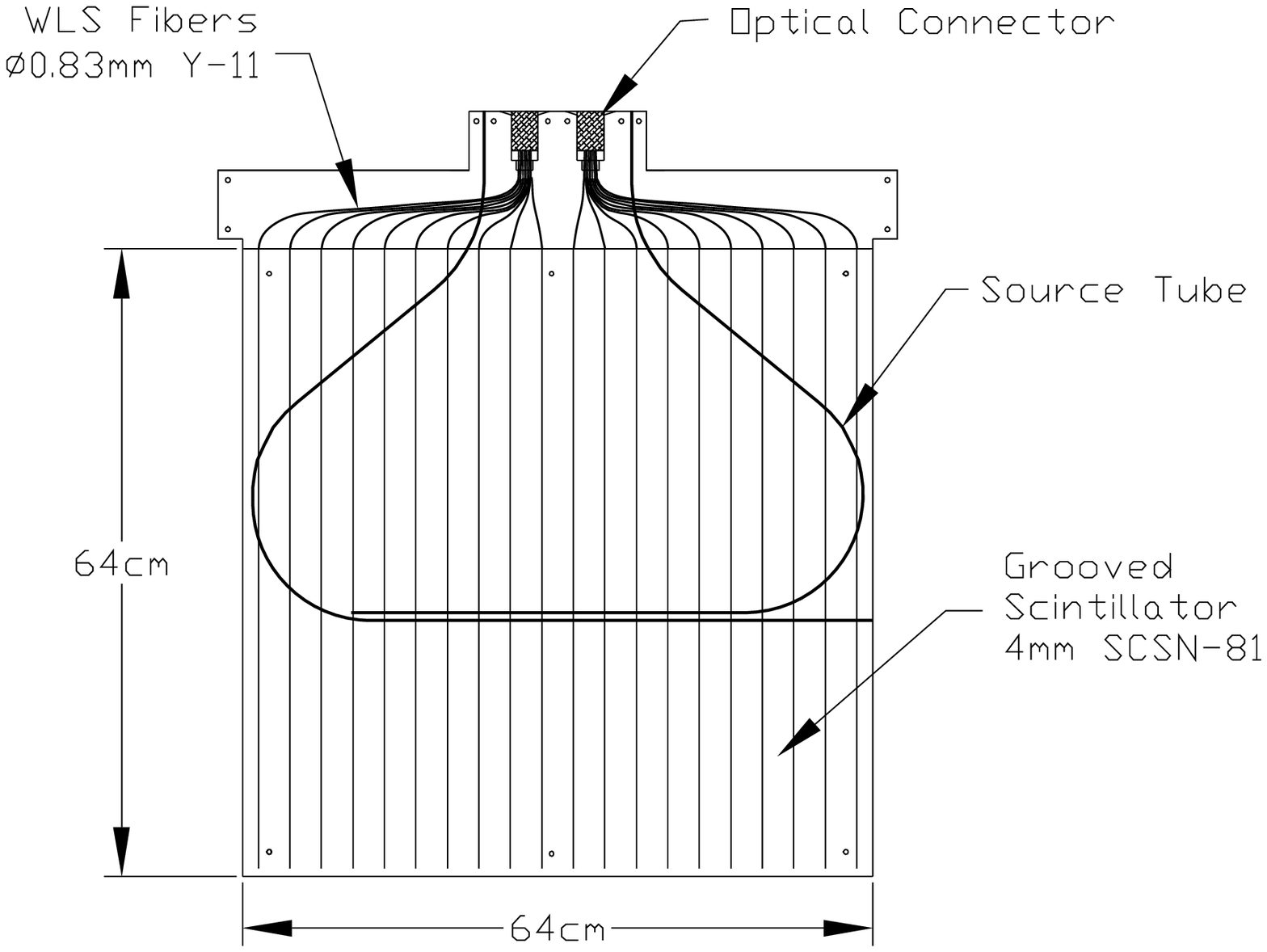}}
\caption{Design of the 64~cm $\times$ 64~cm scintillator
tile for the 1995 CMS Test Beam HCAL module.}
\label{tile95}
\end{center}
\vspace{-0.3cm}
\end{figure}

For the H2(1995) and H4(1995) tests, each scintillator tile
is read out by 20 parallel WLS fibers. 
Figure~\ref{tile95} shows the details of the design of
a single scintillator  tile.
Twenty parallel grooves, spaced every 3.2~cm,  are machined with a
ball-groove tool bit
in 4~mm thick SCSN-81 scintillator plates.
The tiles are painted at the edges using Bicron BC-620 white reflective
paint, wrapped with white reflective
$Tyvek$~\cite{tyvek}, and sealed for light leaks with a black 
$Tedlar$~\cite{tedlar}.
The WLS fibers are mirrored at one end, placed in the grooves, and  
epoxied in groups of ten into two optical connectors placed at the
edge of each scintillator tile, as shown in Figure~\ref{tile95}.
The
stainless steel tubes
are installed inside a 2~mm plastic cover and used to guide 
a radioactive \Cs source  for calibration purposes.
The entire package is held together with a set of brass rivets.

\begin{figure}
\begin{center}
\epsfxsize=4.5in
\mbox{\epsffile{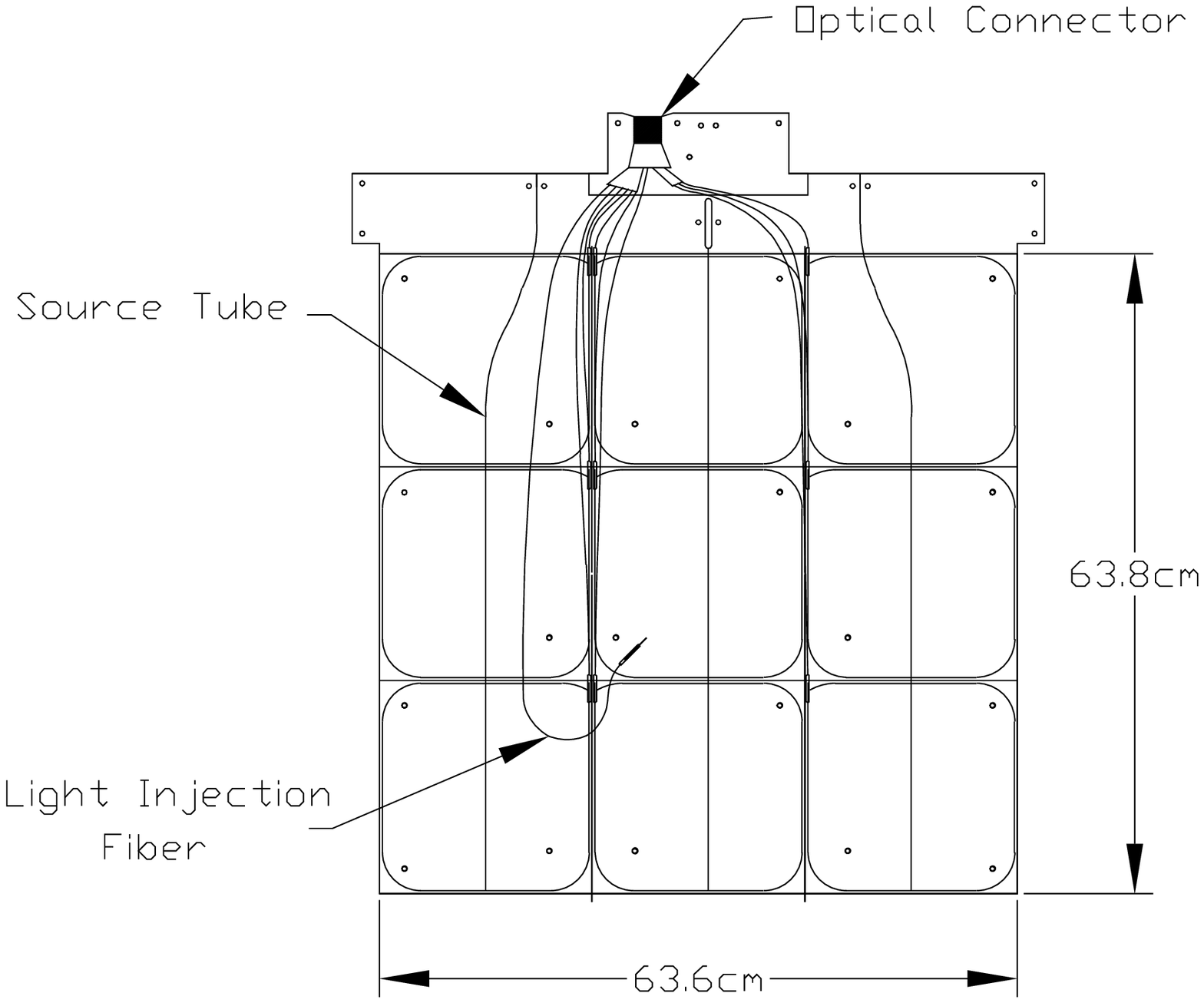}}
\caption{Design of the 64~cm $\times$ 64~cm scintillator
 tile for the 1996 CMS Test Beam HCAL module.}
\label{tile96b}
\end{center}
\vspace{-0.3cm}
\end{figure}

\begin{figure}
\begin{center}
\epsfxsize=4.5in
\mbox{\epsffile{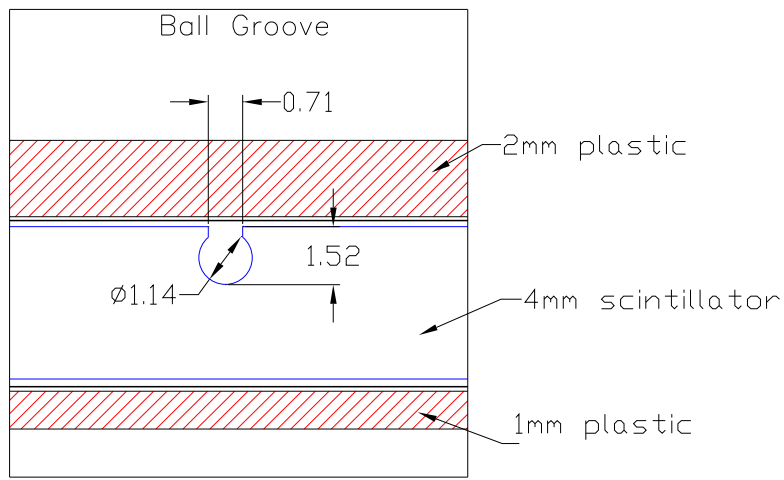}}
\caption{Cross sectional view of the scintillator tile for the CMS
Test Beam HCAL module. The dimensions in the drawing are shown in mm.}
\label{tile96b_cross}
\end{center}
\vspace{-0.3cm}
\end{figure}

For the H2(1996) tests, a different set of scintillator tiles
is used. These tiles are closer
to the tower geometry of the final CMS design.
Each tile is segmented into a 3$\times$3 matrix
read out by nine WLS fibers inserted into  $\sigma$ pattern grooves and
is connected to a single PMT.
The scintillator tile layout is shown in Figure~\ref{tile96b}. 
The tiles are made from 4~mm thick SCSN-81 scintillator produced by
Kuraray. Multiclad Y-11 WLS fibers, made by Kuraray, 0.83~mm in diameter,
are used. The cross-sectional view of a scintillator tile is shown in 
Figure~\ref{tile96b_cross}. Scintillator tiles are packaged between
2~mm thick and 1~mm thick plastic covers. 
The transverse uniformity of response of the scintillator plates,
measured~\cite{qctb95} using a radioactive photon source, has an rms of 4\%.

\begin{figure}
\begin{center}
\epsfxsize=4.5in
\mbox{\epsffile{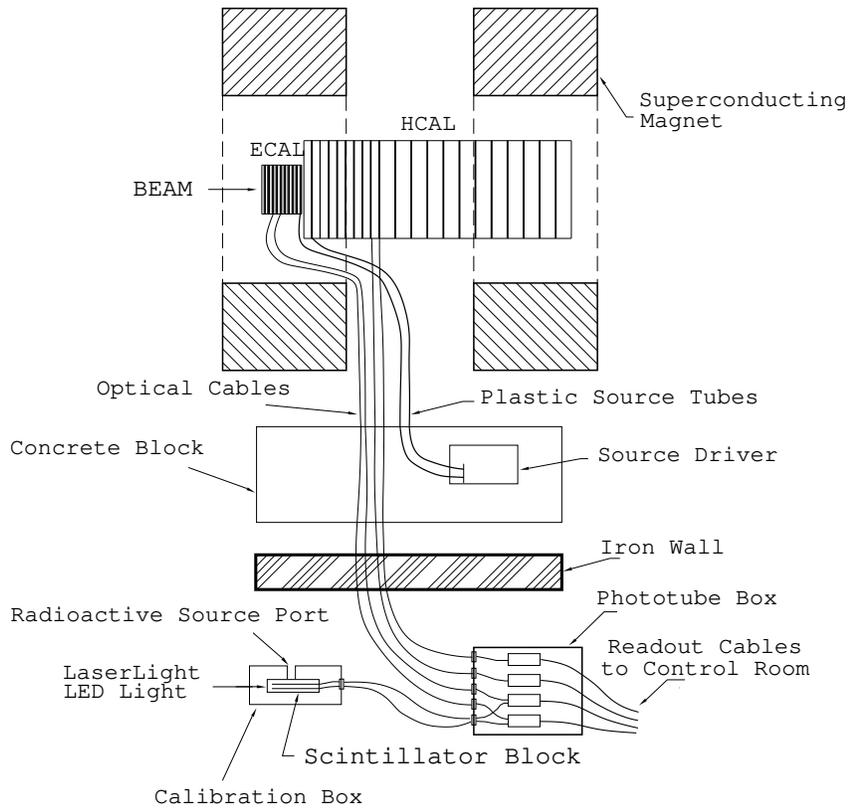}}
\caption{Top view of the CMS test beam site at CERN H2 beamline.
The drawing shows various calibration sub-systems.}
\label{testb2}
\end{center}
\end{figure}

A schematic view of the CERN H2 test beam setup is shown in
Figure~\ref{testb2}. The calorimeter is placed inside a superconducting
magnet, which provides
a magnetic field of up to 3 Tesla. 
Each scintillator tile is read out
independently by a 10 stage, EMI 9902 KA
photomultiplier tube.
Optical cables carry light from the scintillator tiles to the PMT box.
In order to avoid large PMT gain variations from the fringe
fields of the magnet, the PMTs are located
approximately 5~m away from the magnet,
behind a 1~m thick iron wall. 
At this location, the fringe
field for the magnet at 3~Tesla is approximately 50 Gauss.
Therefore, each PMT can be adequately shielded by a
regular inner $\mu$-metal shield and  an outer, 3~mm thick,
soft 
iron pipe. 

\begin{figure}
\begin{center}
\epsfxsize=4.5in
\mbox{\epsffile{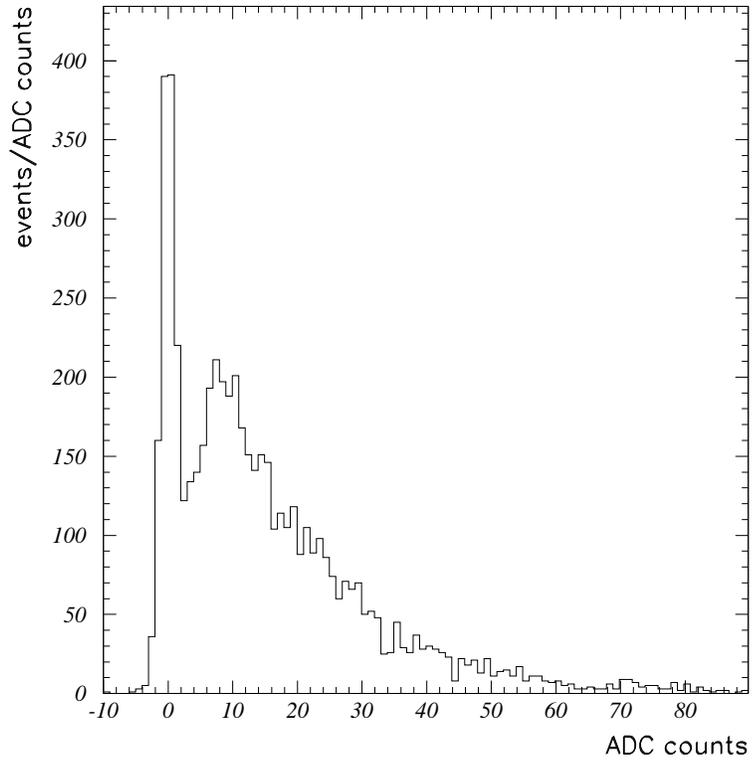}}
\caption{H2(1995) data: 
ADC spectrum of 225 GeV/c muons in a single sampling
scintillator layer.
Based on the observed inefficiency of the  layer of approximately 20\%,
the average number of photoelectrons per minimum ionizing
particle per layer is  1.6 pe/mip.
}
\label{tb96_muons_t12}
\end{center}
\end{figure}

The phototube calibration system is a
crucial part of the setup. 
In order to understand the effects
of the B field on the phototube gains,
several redundant calibration schemes to monitor the PMT gain variation
are used.
The relative calibration of the scintillator tiles is established with
an accuracy of $\approx$ 3\%
by equalizing the average response of each layer to muons. 
Figure~\ref{tb96_muons_t12} shows an ADC spectrum for 225 GeV/c muons
 traversing  a single scintillator layer.
The absolute light yield of each scintillator tile, determined using
muons,
is approximately 1.5 photoelectrons per minimum ionizing particle.
This light yield is sufficiently high such as to
not contribute significantly~\cite{bodek}
to the hadronic energy resolution of HCAL.

An independent method
of  monitoring the gain of each tile is provided by a system of 
radioactive source guide tubes~\cite{virgil}.
A radioactive  \Cs source located at the tip of 
a long stainless steel wire is periodically
inserted into guide tubes embedded in 
the plastic cover sheet for each tile.
The source is  moved using a computer controlled source driver.

The muon calibration and the wire source calibration schemes 
monitor variations in the product of the light yield 
of the scintillator tiles
and the gain of the photomultipliers. 
In order to separately
monitor variations in phototube gains 
(from temperature dependence and B fringe field effects),
each PMT is also connected 
to a special scintillator block
via a set of separate Y11 calibration fibers.
The scintillator block is excited using 
either 
a laser beam,  or a \Cs source. 
In addition, the Y11 calibration fibers can be excited by a light 
from a blue light emitting diode (LED).
The scintillator
block is located in a region
with a low (approximately 50 Gauss)  magnetic field, and thus is in a
region
where scintillator brightening
from the fringe magnetic field is below 1\%~\cite{desy1}.

\section{Calorimeter Performance in a Magnetic Field}

  The CMS hadron calorimeter, which uses plastic scintillators,
  will operate in a high magnetic field (4 Tesla).
  As has been previously
  observed~\cite{desy1,desy2,bfield}, the light yield of plastic
  scintillators 
  placed in high magnetic fields increases by 5 to 8\%. This increase
  is due to polymer
  excitation that increases the energy of the short wavelength
  primary light in the ultra-violet (UV) region.

In addition, the presence of a magnetic field can  
affect~\cite{kryshkin,abramov} the energy deposited in active parts of a
sampling calorimeter. To understand this, the response of HCAL to muons,
electrons, and pions has been studied as a function of magnetic
field strength and orientation for several different sampling
configurations.  For all magnetic field studies the ECAL module was
moved out of the beamline.
During the H2(1995) tests, the B-field was oriented perpendicular
to the scintillator planes, corresponding to the endcap HCAL
configuration, while during the H2(1996) tests, the B-field 
was parallel to scintillator
planes, corresponding to the 
barrel HCAL configuration.

\subsection{B field perpendicular to the scintillator planes 
 (endcap configuration)}

\begin{figure}
\begin{center}
\epsfysize=4.5in
\mbox{\epsffile{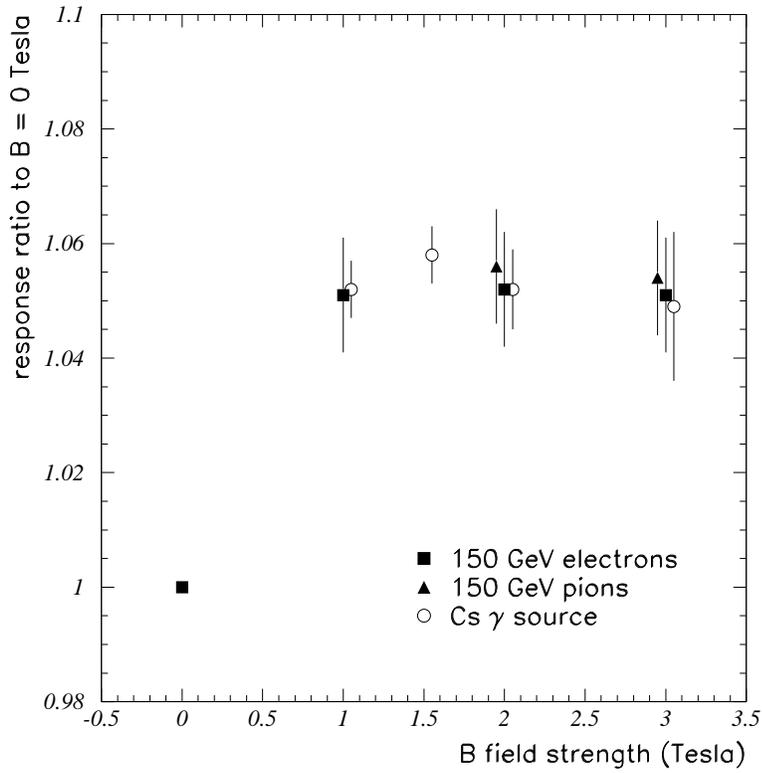}}
\caption{H2(1995) data: average energy response
of the tile/fiber calorimeter to pions and electrons
as a function of B field, relative to response for B = 0 Tesla field.
Also shown on the plot is the relative response of scintillator to  
radioactive gamma ray  calibration source, as a function of B field.
Here the B field lines are
perpendicular to the scintillator plates (endcap configuration).
}
\label{cms_95_barrel_em_had_sc_bw}
\end{center}
\vspace{-0.3cm}
\end{figure}

  When the magnetic field lines are perpendicular to the scintillator
  planes, the intrinsic
  light yield of the scintillator is higher than 
  in the case where there is no magnetic field. This scintillator brightening
  effect leads to 
  the same overall increased response of the calorimeter to muons,
  electrons, and pions. 
  Figure~\ref{cms_95_barrel_em_had_sc_bw} shows the response
  of HCAL to pions and electrons as a function of B field
  relative to B=0 Tesla.
  The observed $\approx$ 5\% increase
  in the response of the calorimeter
  to particles is consistent
  with the increased light yield of the scintillator plastic,
  as measured by the calibration system using the radioactive \Cs 
  photon source.

\begin{figure}
\begin{center}
\epsfysize=4.5in
\mbox{\epsffile{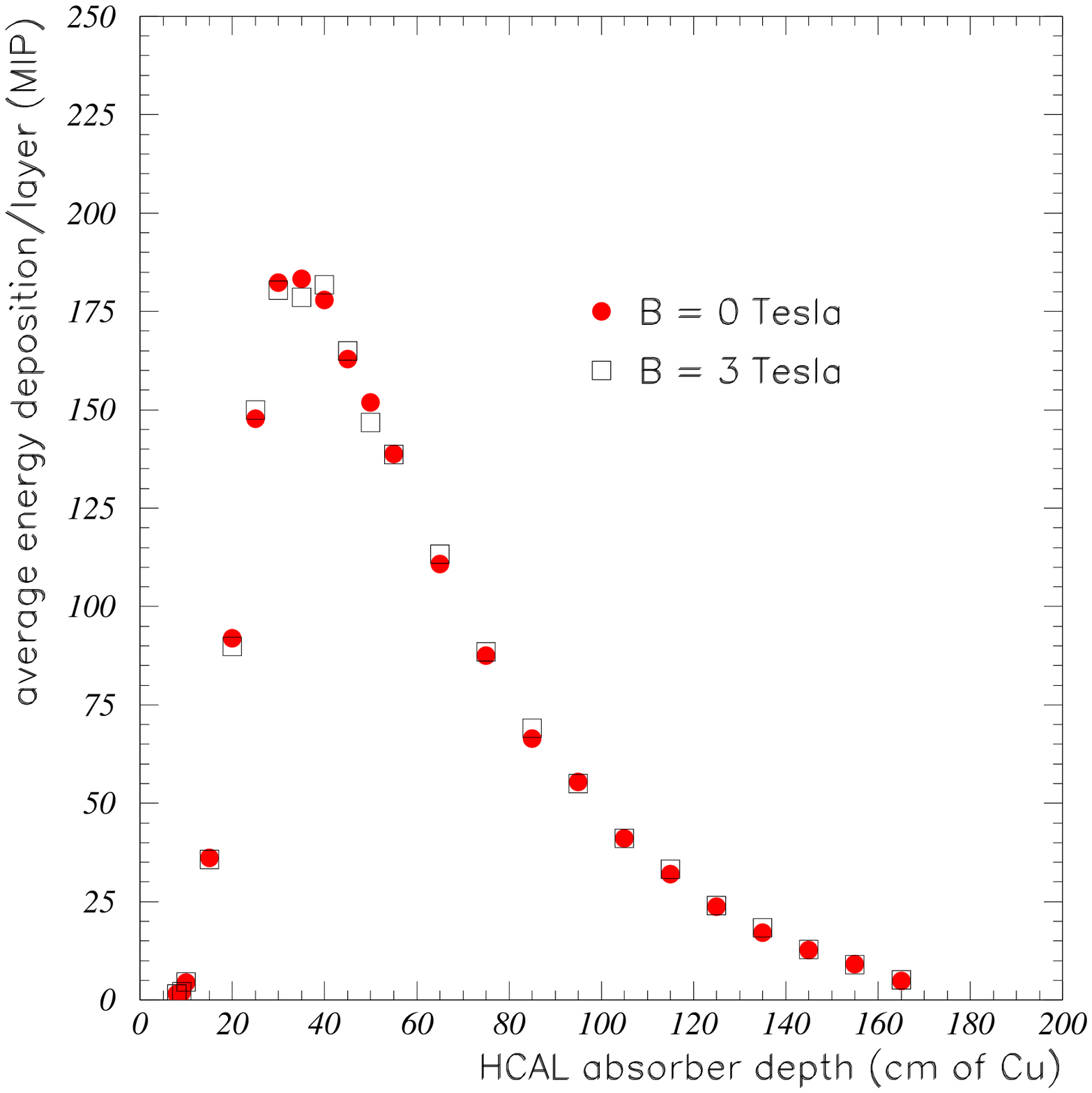}}
\caption{H2(1995) data:
comparison of 300 GeV/c pion shower profiles for B=0 and B=3 Tesla
magnetic fields. The B field lines 
are perpendicular  to the scintillator plates (endcap configuration).
The pion shower profiles are divided by
the average muon
response for each layer, which corrects for the
overall scintillator brightening effect.
}
\label{h2-1995-p300-vs-B}
\end{center}
\end{figure}

  Figure~\ref{h2-1995-p300-vs-B} shows a comparison of 
  the average shower profile for
  300 GeV/c pions
  for two magnetic field settings: B=0 Tesla and
  B= 3 Tesla. 
  The average energy deposition per sampling layer, in
  minimum ionizing particles (mips),
  is plotted as a function of HCAL absorber depth.    
  Since the gain of each scintillator layer  
  is established by equalizing its response to
  muons independently  for the  0~T and 3~T data sets, 
  the  effect of scintillator brightening
  is already removed.
  As can be seen in the figure, 
  the shape of the pion shower profiles remains unchanged  
 in the presence of a 3 Tesla magnetic field perpendicular
  to the scintillator planes.
  
  In summary, we conclude that a 
  B field perpendicular to the scintillator planes
  results in an overall increase in the scintillator light yield.
  This scintillator brightening effect can be effectively
  measured by the calibration system using a radioactive $\gamma$ source.

\subsection{B field parallel to the scintillator planes
 (barrel configuration)}

\begin{figure}
\begin{center}
\epsfysize=4.5in
\mbox{\epsffile{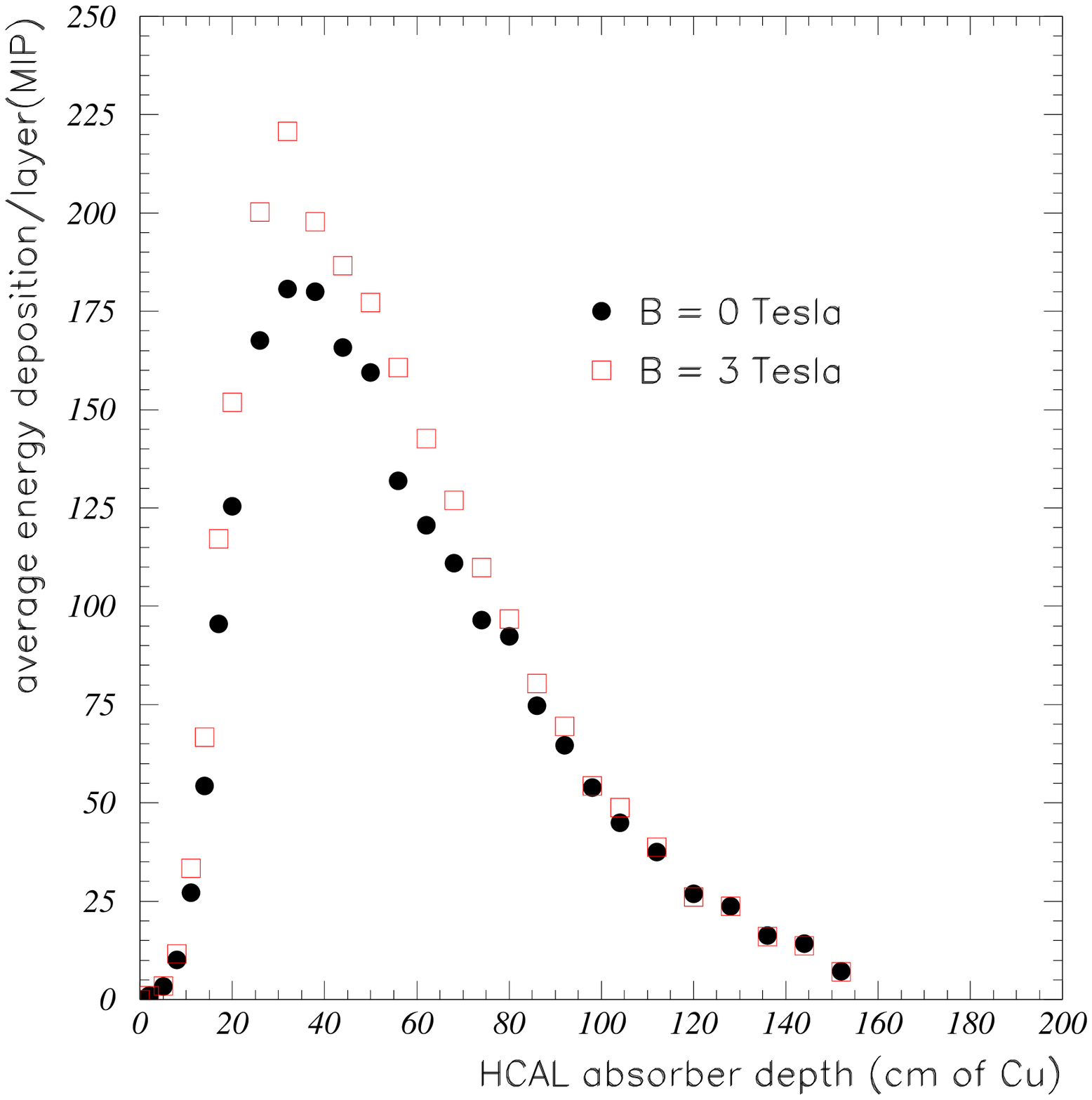}}
\caption{H2(1996) data:
comparison of 300 GeV/c pion shower profiles for B=0 and B=3 Tesla
magnetic field. Here,  the B field lines 
are parallel to the scintillator plates (barrel configuration).
The pion shower profiles are divided by
the average muon
response for each layer, which corrects  for the
overall scintillator brightening effect.
}
\label{h2-1996-p050-vs-B}
\end{center}
\end{figure}

  In contrast, when the magnetic field lines are parallel to the
  scintillator planes, the observed
  average hadron shower profiles
  are altered.
  A comparison of the
  average observed shower profiles for 300/c GeV pions 
  as a function of the HCAL absorber depth, for B = 0 and B = 3 Tesla,
  is shown in Figure~\ref{h2-1996-p050-vs-B}.
  Energy deposited by hadron
  showers in mips shows
  an additional increased response, in the B = 3 Tesla field, relative to
the B= 0 Tesla case.

\begin{figure}
\begin{center}
\epsfysize=4.5in
\mbox{\epsffile{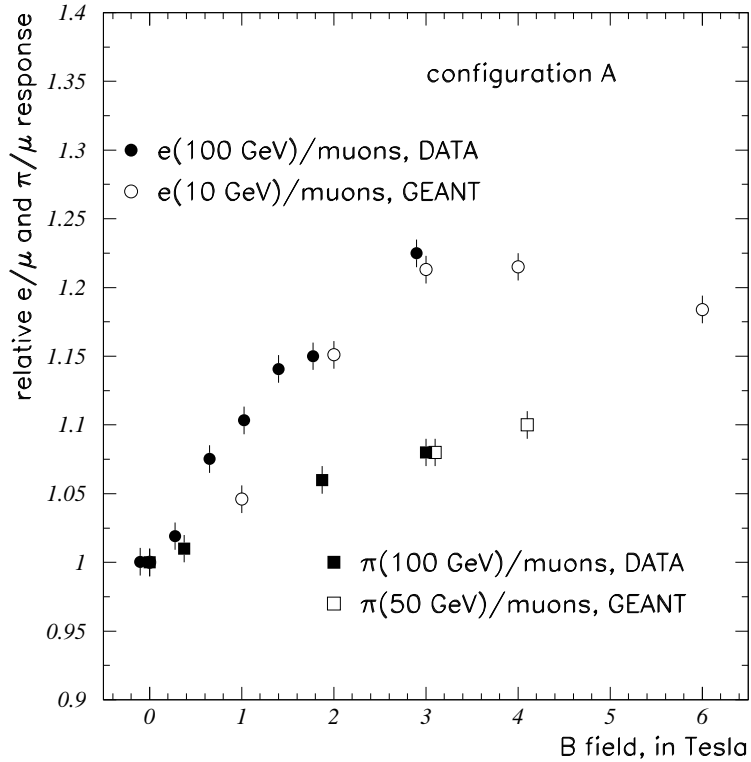}}
\caption{H2(1996) data: effect of a B field on the average energy response
of the tile/fiber calorimeter to pions and electrons
divided by the average muon response. 
Response ratio is normalized to 1 for B= 0 Tesla. 
Also shown are the results
from a GEANT simulation. Here, the B field lines are
parallel to the scintillator plates (barrel configuration). 
The position of the scintillator package 
relative to the incident  beam direction is the following:
1~mm plastic + 4~mm scintillator + 2~mm plastic (configuration A).
The overall scintillator brightening B field effect is
removed since the response of electrons and pions are
divided by the average muon response in each layer.
}
\label{e-vs-B-geant-98a}
\end{center}
\vspace{-0.3cm}
\end{figure}

  The relative average response of HCAL to
  100 GeV/c electrons and pions as a function of magnetic  field 
  is shown in Figure~\ref{e-vs-B-geant-98a}.  
  As indicated by the data,
  the HCAL response to electrons in mips
  increases by approximately 20\%  at B=3 Tesla, relative to B=0 Tesla.
  However, the response for pions in mips
  increases by only $\approx$ 8\%.
  The results
  are compared to simulations
  using the GEANT~\cite{ref_geant} Monte Carlo program 
 (10 GeV/c electrons and 50
GeV/c pions). 
As shown in the figure, the simulations reproduce data well.

  Studies using a GEANT Monte Carlo simulation
  indicate that the increased response is due to a geometric effect
  resulting from a change in the  path length of low energy electrons 
  (between 1 and 10 MeV) exiting the absorber plates and
  traversing a circular path
  in the scintillator layer (because of the strong magnetic field).
  The  radius of curvature of a
  1 MeV/c electron in a 3 Tesla magnetic field
  is approximately 1~mm. The
  size of this  geometric effect is expected to be 
  proportional to the strength of the B field and
  depends on the detailed structure~\cite{abramov} and composition
  of the  absorber/scintillator package.
 
\begin{figure}
\begin{center}
\epsfysize=4.5in
\mbox{\epsffile{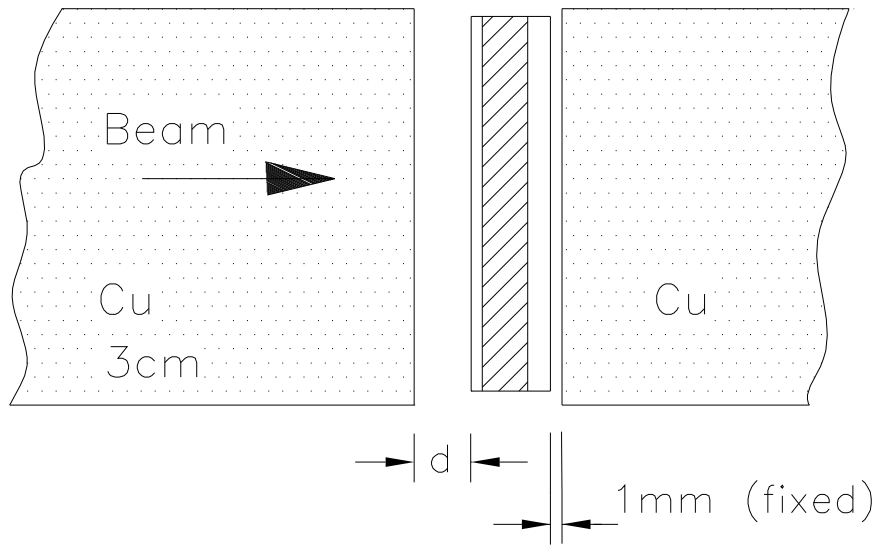}}
\caption{
Orientation of the scintillator package with respect to the  
beam direction inside the absorber gap (configuration A):
1~mm plastic + 4~mm scintillator + 2~mm plastic.
The beam is incident
from the left.
In configuration B, the orientation of the package is reversed:
2~mm plastic + 4~mm scintillator + 1~mm plastic.
}
\label{scint_orient}
\end{center}
\vspace{-0.3cm}
\end{figure}

\begin{figure}
\begin{center}
\epsfysize=4.5in
\mbox{\epsffile{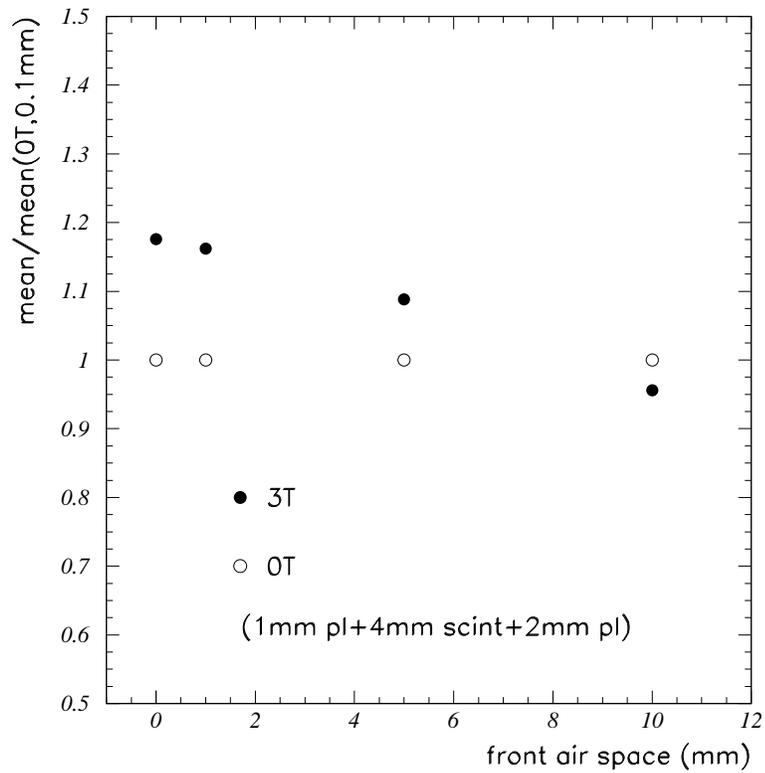}}
\caption{ Results of
a GEANT simulation of the HCAL calorimeter response to 10 GeV electrons
(in a 3 Tesla magnetic field parallel to the scintillator
plates) as a function of the air gap (d) between the scintillator package and
the most upstream absorber plate for configuration A.
}
\label{mag_geant_sim}
\end{center}
\vspace{-0.3cm}
\end{figure}

  A GEANT study models the calorimeter response to 10 GeV
  electrons  as a function of the air gap between the scintillator 
  package and the upstream absorber (see Figure~\ref{scint_orient}) 
  in the presence of a strong magnetic field. The results are 
  summarized in Figure~\ref{mag_geant_sim}.
  The increase in the electron response is reduced as the distance between
  scintillator package and upstream absorber is increased.

   The results presented so far have been taken with 
 the scintillator package placed in the absorber gap 
  in the following way:
  1~mm plastic + 4~mm~scintillator + 2~mm plastic,
  with respect to the beam direction.
  In  this configuration (configuration A), 
the  scintillator is separated from the upstream-most absorber plate 
by 1~mm $\pm$ 0.5~mm of air and 1~mm of plastic.
  The Monte Carlo studies indicated that the effect of a parallel
  magnetic field depends on
  the distance between the scintillator and the most upstream absorber.
  Therefore, we have investigated the calorimeter response to pions,
  electrons, and muons 
  with  the reversed orientation of the scintillator package
  in the absorber gap (configuration B).
  In configuration B, the position  of the scintillator package
  relative to the beam direction is:  2~mm plastic + 4~mm scintillator + 
  1~mm plastic. In the case of configuration B, 
the  scintillator is separated from the upstream most absorber plate 
by 1~mm $\pm$ 0.5~mm of air and 2~mm of plastic.

\begin{figure}
\begin{center}
\epsfysize=4.5in
\mbox{\epsffile{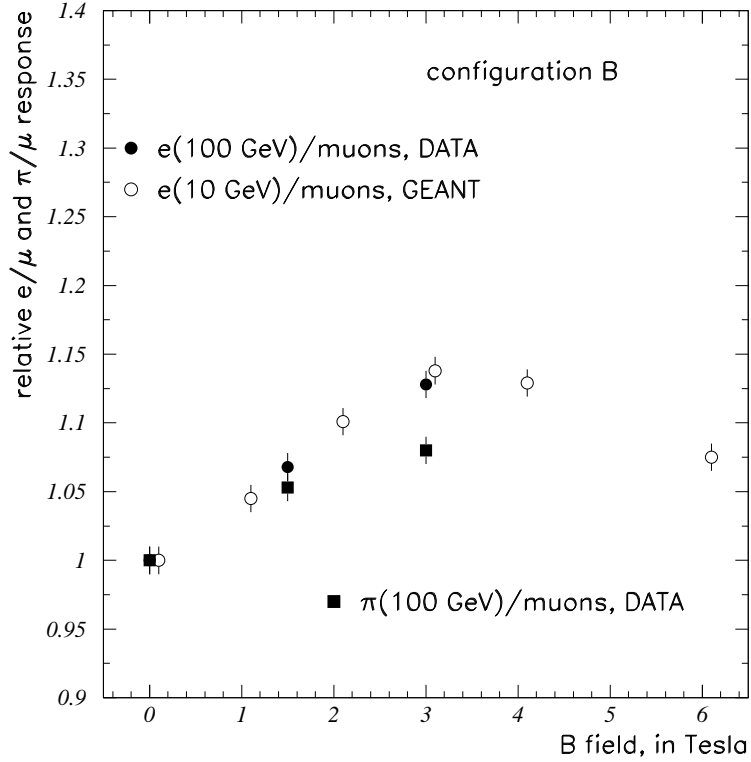}}
\caption{H2(1996) data:
effect of a B field on the average energy response 
of the tile/fiber calorimeter to electrons and pions.
Response ratio is normalized to 1 for B= 0 Tesla. 
Also shown are results from GEANT simulation.
Here, the B field lines are
parallel to the scintillator plates (barrel configuration). 
The position of the scintillator package 
relative to the incident beam direction
is as follows:
2~mm plastic + 4~mm scintillator + 1~mm plastic (configuration B).
The
overall scintillator brightening B field effect is
removed since the response for electrons and pions is
divided by the average muon response.
}
\label{e-vs-B-geant-98b}
\end{center}
\vspace{-0.3cm}
\end{figure}

  Figure~\ref{e-vs-B-geant-98b} shows 
  the average electron and pion response of  HCAL in mips as a function
of B field,
  relative to the response at B= 0 Tesla, 
  measured with the configuration B scintillator package.
   Here, we measure the response of electrons at B = 3 Tesla to be  only
   14 $\pm$ 1\% higher (relative to B =0 Tesla), compared to 20\% $\pm$ 
   1\%
   effect
   for configuration A. GEANT simulation reproduces the electron  data
   well. The additional 1~mm plastic upstream of the scintillator in
   configuration B (vs configuration A) helps range out more low energy
   electrons coming out of the upstream absorber plate.

  In order to minimize the effect of the magnetic field 
  on the response of the CMS hadron  barrel calorimeter,
  one would like to maximize the distance between the scintillator
  and the upstream absorber.
  Therefore 
  the configuration B
  of scintillator placement in the gaps between absorber
  plates has been chosen for the CMS HCAL design. 
  A system of thin brass (Venetian Blind type) springs
  pushes the scintillator tiles radially
  outwards, such that the scintillator is  always positioned closest to
  the downstream absorber plate.
  The use of springs ensures
  that the distance between inner absorber and the scintillator plate will
  consist of 2~mm $\pm$ 0.5~mm of air and 2~mm of plastic.   
  
\begin{figure}
\begin{center}
\epsfysize=4.5in
\mbox{\epsffile{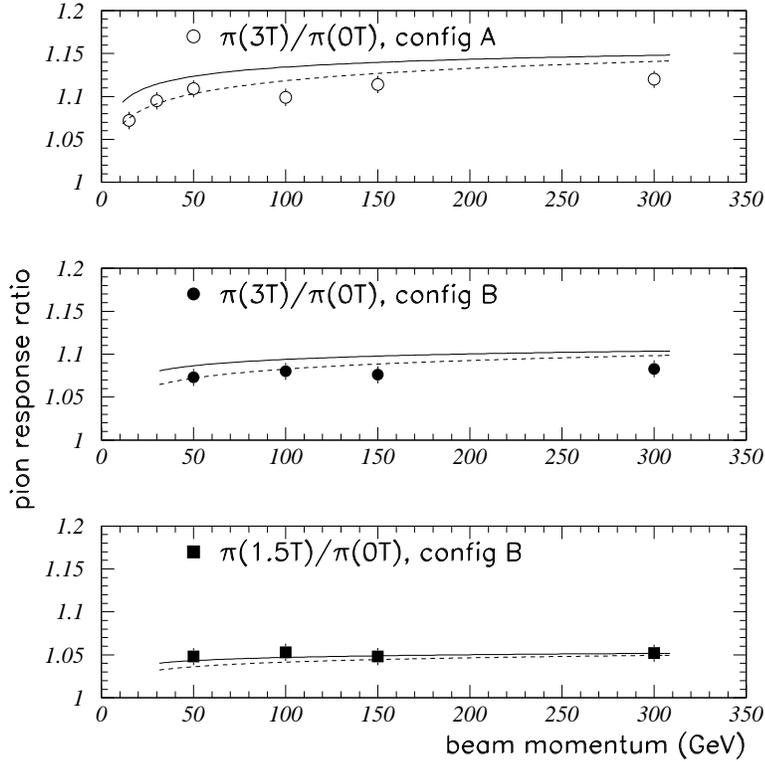}}
\caption{H2(1996) data: the
ratio of the pion response of HCAL at 3 Tesla and 1.5 Tesla to
the pion  response at 0
Tesla versus beam energy. Here, the B field lines are
parallel
to the scintillator
plates (barrel configuration). 
The data is shown for two configurations of scintillator package inside
absorber gap:
1~mm plastic + 4~mm scintillator + 2~mm plastic (configuration A) and
2~mm plastic + 4~mm scintillator + 1~mm plastic (configuration B).
The  lines correspond to the predictions using 
Groom (solid lines) and Wigmans (dashed lines)  parametrizations of the
fraction of the electromagnetic
component in hadronic showers.
}
\label{pi3_pi0}
\end{center}
\vspace{-0.3cm}
\end{figure}

  Figure~\ref{pi3_pi0} shows the
  ratio of HCAL response to pions as a function of pion energy, 
  at 3 Tesla and 1.5 Tesla, relative to its
response at 0 Tesla for the two configurations of the scintillator package
inside the absorber gap.
  The lines shown  on the Figure~\ref{pi3_pi0} indicate  the
  values for $\pi(3T)/\pi(0T)$ and $\pi(1.5T)/\pi(0T)$ 
  as a function of particle
  momentum. These functions were calculated  using the 
  Wigmans~\cite{wigmans}
and Groom~\cite{groom} parameterizations\footnote{Detailed 
  study of 
  the average fraction of the electromagnetic component in pion showers,
  $F(\pi_{0})$, 
  and extraction of $e/h$ 
  is presented in section 6.}  of
  the average fraction of the electromagnetic component in pion showers.
    The average fraction of the electromagnetic component in pion showers 
  increases as a function of pion energy. With $e/h$ of HCAL greater 
  than unity,
  this increase results in non-linearity of HCAL response to pions. For 
example, the relative response of the
  calorimeter to 300 GeV/c pions is higher by 
$\approx$ 5\%, 
with respect to 50 GeV/c pions.
  However, a 3 Tesla magnetic field increases the $e/h$ ratio for HCAL 
  by $\approx$  20 \%  for configuration A and 14\% for configuration B,
  see  Figs.~\ref{e-vs-B-geant-98a} and~\ref{e-vs-B-geant-98b}.
  Therefore, a 3 Tesla magnetic field is expected to 
  increase the non-linearity 
  of HCAL's response to pions between 50 GeV/c  and
  300 GeV/c by an additional $\approx$  2-3\%.

  The above estimate is relevant only for the subset of
  single pions which interact in HCAL (minimum ionizing in ECAL).   
  However, in the actual CMS detector, the common situation is that
  of a jet of particles depositing a large fraction of energy in the
  ECAL calorimeter. Since the ECAL calorimeter is not sensitive to 
  magnetic field effects, the additional non-linearity from the
  3 Tesla B field is greatly reduced.

  In summary, a B field parallel to the scintillator planes results in 
  an increased response due to two effects: (a) the
  scintillator brightening effect and (b) an additional increase in the
  response to the electromagnetic component of
  hadron showers (geometrical effect). The latter effect
  is strongly dependent on the placement of the scintillator packages 
  with respect to the absorber plates of the calorimeter.  

  The results reported in this paper are consistent with earlier reports
  on the behavior of plastic scintillators in magnetic
fields~\cite{desy1,bfield} and on the 
  influence of magnetic fields on response of scintillator based
  calorimeter~\cite{desy2}. Note however, that in the latter publication
  the conclusion that electron response follows the dependence of the
  light yield ("brightening effect") is true only for magnetic fields 
  below 0.3 Tesla. As shown in this report, in the presense of parallel
magnetic fields above 0.3 Tesla, the geometric effect leads to
additional increase in response of the sampling calorimeters to electrons. 

  We also note that the origin of the latter  effect is  due
  to the relative geometry of the absorber
  with respect to the active elements of the calorimeter and is not sensitive
  to any special property of
  the scintillator as the active sampling medium.
  Therefore,  similar effects are expected for 
  any type of sampling calorimeters, with similar magnetic field/absorber
  configurations, independent of the choice of material
  for the active sampling medium.
\section{Performance the HCAL calorimeter with a Pb-Scintillator ECAL}

During the H2(1995) running period, a Pb-scintillator  ECAL calorimeter
was placed in front of the HCAL calorimeter. Linearity of the response
and energy resolution of the combined system was studied for electrons,
pions, and muons with momenta from 50 to 300 GeV/c. 

In the following
discussion we consider two different samples of pion data. The first,
which we refer to as "mip-in-ECAL" pions,
includes only pions which are minimum ionizing in 
ECAL, i.e. do not interact in ECAL and begin  interacting in HCAL.
This sample is selected by
requiring that the energy deposition in ECAL is consistent with
minimum ionizing deposition, i.e. is less than 2.5 GeV.
The second sample, which we refer to as the "full pion sample", includes
all pions, i.e. interacting in either ECAL or HCAL. 

\begin{figure}
\begin{center}
\epsfxsize=4.5in
\mbox{\epsffile{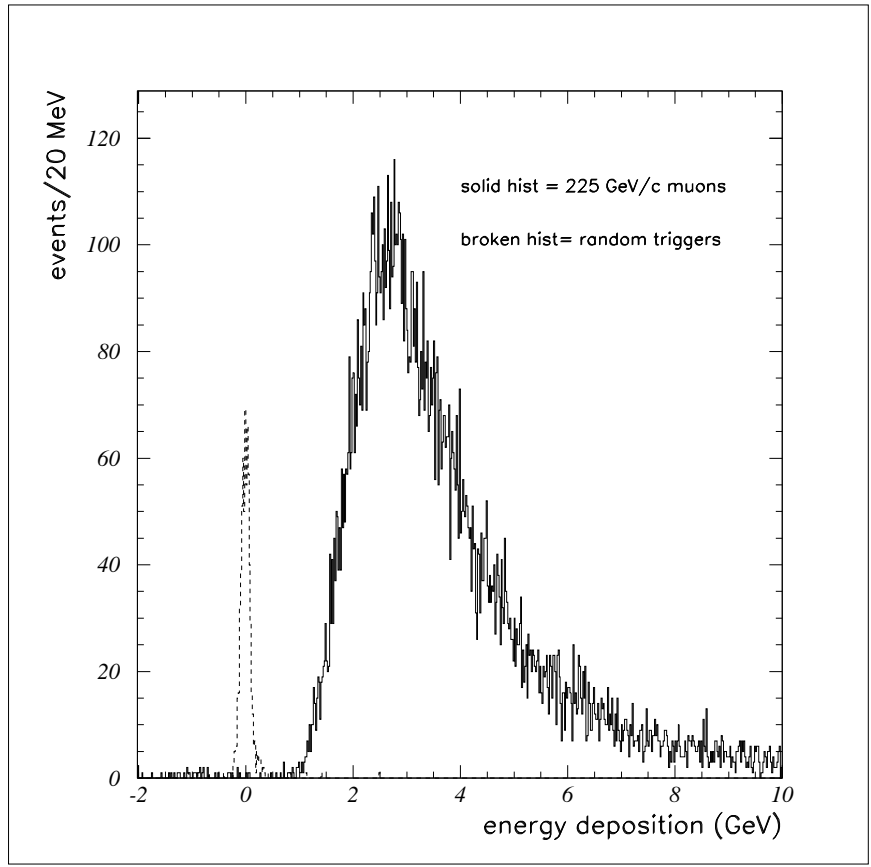}}
\caption{H2(1995) data: 
energy deposited by 225 GeV/c muons in HCAL.
The dashed line shows the energy reconstructed in HCAL for
random triggers (pedestal events). The pedestal peak
has a rms width of 80 MeV of equivalent hadron energy.
}
\label{tb96_muons_ped}
\end{center}
\end{figure}

The absolute energy scale of the HCAL calorimeter 
is set using 300 GeV/c "mip-in-ECAL" pions.
The average pulse height deposited by muons in HCAL
corresponds to approximately 3 GeV of equivalent hadronic energy.
Figure~\ref{tb96_muons_ped} shows the  energy deposited by
225 GeV/c muons in HCAL. 
A pedestal trigger is a random trigger during the beam.
This pedestal distribution is shown
as a dashed line in Fig.~\ref{tb96_muons_ped}.
The width of the energy distribution for random pedestal triggers is
equivalent to 80 MeV. Hence, 
the contribution of electronic noise to the 
energy resolution of HCAL is negligible.

The absolute energy scale of the ECAL calorimeter 
is set using 50 GeV/c electrons.
The energy response of the Pb-scintillator sampling
ECAL calorimeter  ($\approx$ 2.9 X$_0$ sampling)
for electrons with momenta in range  50 to 150 GeV/c 
is linear to within 1\%. The fractional energy resolution for 50 and 
100 GeV/c electrons is 7.2\% and 6.2\%, respectively.

The energy response 
of the combined Pb-scintillator ECAL+HCAL calorimeter
system is calculated as follows:
\begin{equation}
	E_{TOT} = \sum_{i}^{ECAL}  \frac{ADC_i*w_i}{\mu_i}/Escale +
                  \sum_{j}^{HCAL}  \frac{ADC_j*w_j}{\mu_j}/Hscale
\end{equation}

Here, $Escale$ and $Hscale$ are constants which define
the absolute energy scales of ECAL and HCAL, and
$ADC_i$ are the ADC values of each
readout layer of ECAL and HCAL.
The relative calibration constants of each layer,
$\mu_i$, correspond to the average muon response
of a layer expressed in units of ADC counts.
The coefficients $w_i$ are proportional to the arithmetic mean of 
the interaction lengths of the absorber plates upstream and downstream  of 
each scintillator layer (the "Simpson" approximation  formula), 
$w = 0.5 \times (t_{upstream} + t_{downstream})$.

\begin{figure}
\begin{center}
\epsfxsize=6in
\mbox{\epsffile{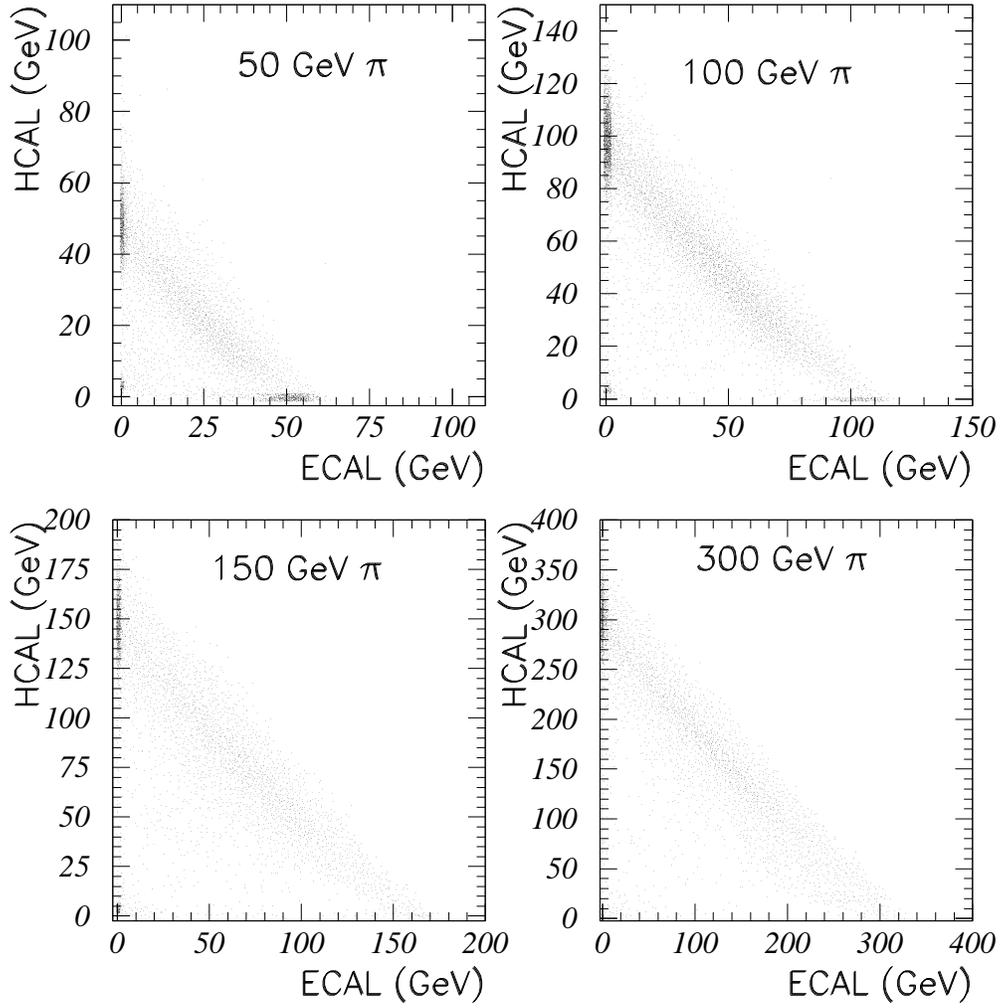}}
\caption{H2(1995) data: scatter plots of $E_{HCAL}$, the energy
deposited in the hadron compartment, (vertical scale) versus 
$E_{ECAL}$, the energy deposited in the Pb-scintillator
 electromagnetic compartment,
(horizontal scale). 
In addition to  pions, there is a contamination
of electrons and muons in the beam.}
\label{h2_scatter}
\end{center}
\end{figure}

Scatter plots of the energies in
 HCAL vs ECAL 
for 50, 100, 150, and 300 GeV/c particles in the hadron beam
are shown in Figure~\ref{h2_scatter}.
As indicated in the figure, the nominal hadron beam, especially 
at the 50 GeV/c and 100 GeV/c beam tunes, contains a
large fraction of electrons.
Pion induced events are selected
by requiring that energy deposited in HCAL is at least 
0.5 GeV.
In order to remove particles that interact upstream
of ECAL, the energy deposition in the
first ECAL layer is required
to be consistent with that of a minimum ionizing particle.

\begin{figure}
\begin{center}
\epsfxsize=6in
\mbox{\epsffile{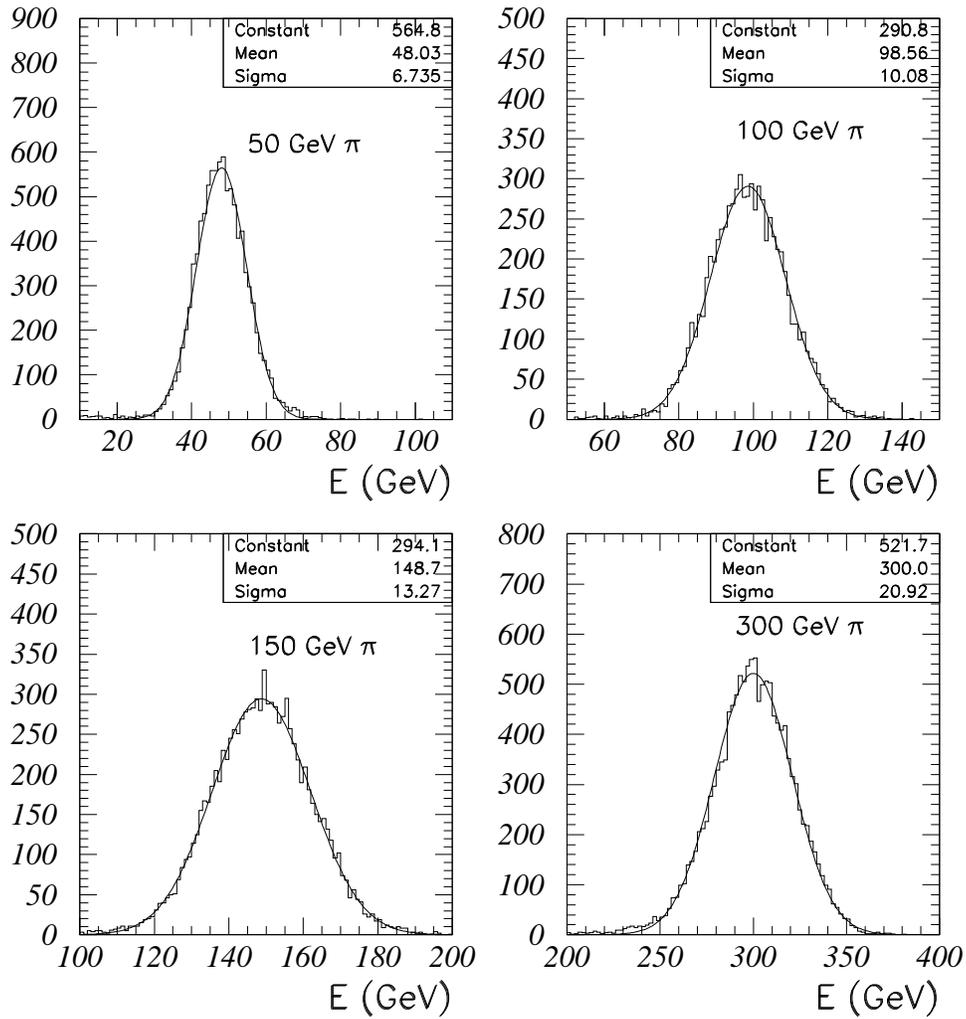}}
\caption{H2(1995) data: reconstructed energy distributions 
for "mip-in-ECAL" pions. 
The minimum ionizing energy deposition in ECAL
has been added to the energy sum. 
}
\label{h2_h_mip}
\end{center}
\end{figure}

\begin{figure}
\begin{center}
\epsfxsize=6in
\mbox{\epsffile{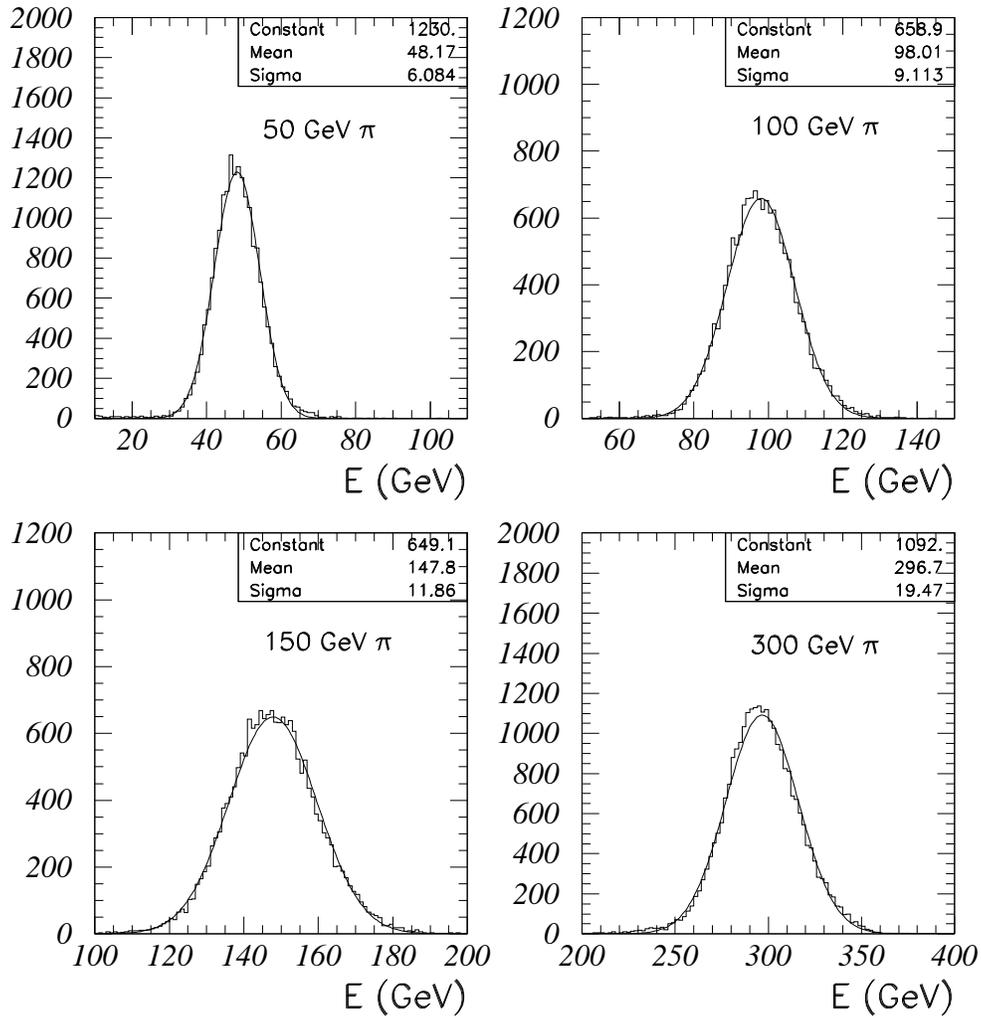}}
\caption{H2(1995) data: distributions of $E_{TOT}$, the sum
of energies
deposited in the hadronic and electromagnetic compartment 
for "full pion
sample".
}
\label{h2_pions_all}
\end{center}
\end{figure}

\begin{figure}
\begin{center}
\epsfxsize=4.5in
\mbox{\epsffile{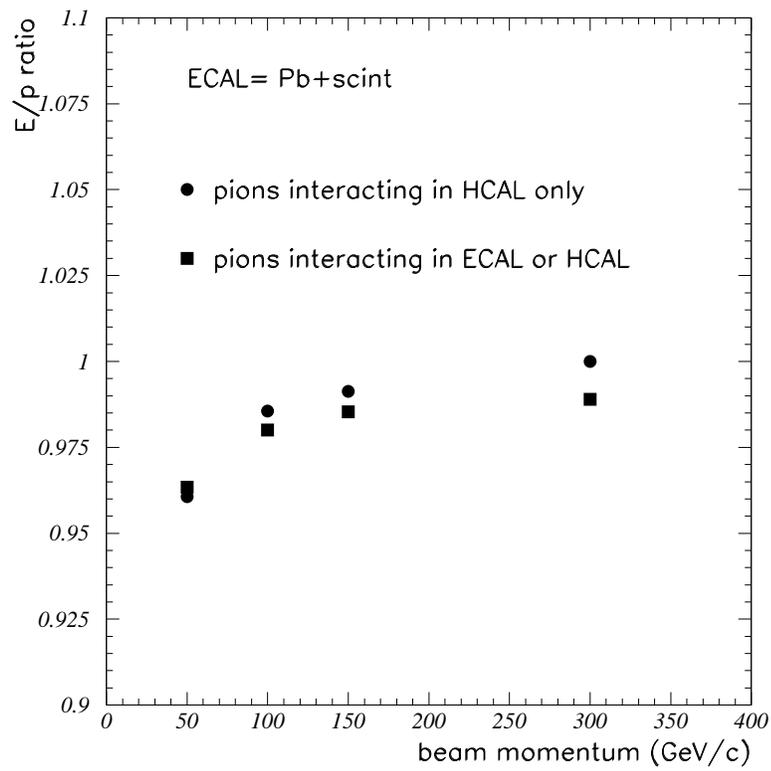}}
\caption{H2(1995) data: linearity of the energy response for
"mip-in-ECAL" pions and "full pion sample".
The statistical errors are smaller than the size of the symbols.}
\label{h2_lin}
\end{center}
\end{figure}

The total (ECAL+HCAL) energy distributions for "mip-in-ECAL"
pions are shown in 
Figure~\ref{h2_h_mip}.
The total energy  distributions for
the "full pion sample" 
are shown in Figure~\ref{h2_pions_all}.
The reconstructed energy distributions are well described by 
Gaussian fits for both sets of data.
Figure~\ref{h2_lin} shows
the linearity versus energy 
for the calorimeter response for "mip-in-ECAL" pions, and
for the "full pion sample" using the  Pb-scintillator ECAL+HCAL
calorimetric system. The average 
response of the combined
Pb-scintillator ECAL+HCAL system  to
the "full pion sample" 
is approximately 1\% lower than the response the calorimeter for the subset
of "mip-in-ECAL" pions.

\begin{figure}
\begin{center}
\epsfxsize=4.5in
\mbox{\epsffile{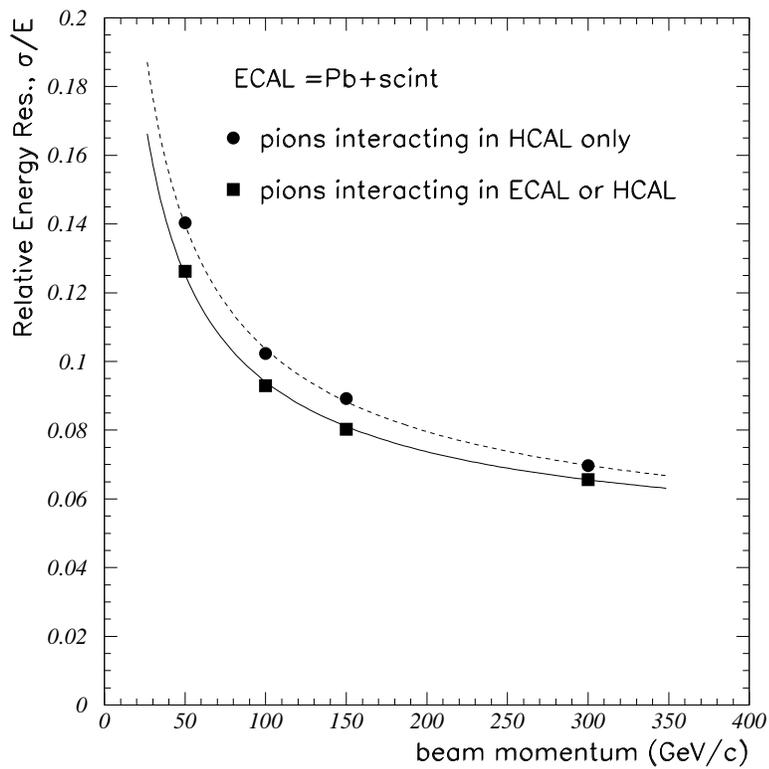}}
\caption{H2(1995) data: fractional energy resolutions for
"mip-in-ECAL" pions and "full pion sample".
The statistical errors are smaller than the size of the symbols.}
\label{h2_res}
\end{center}
\end{figure}

 A comparison of the 
relative energy  resolutions of the calorimeter
for "mip-in-ECAL"  pions and
for the "full pion sample"
is shown in Figure~\ref{h2_res}.
The energy resolutions for these two cases are comparable.
 At low energies,
the energy resolution of the "full pion sample"  
is slightly improved because of the finer sampling of the Pb-scintillator
 ECAL.

The fractional energy resolutions are parameterized by the following
function: 
\begin{equation}
        \sigma_{E}/E = (stoch.~term)/\sqrt{E} \oplus (const.~term)
\end{equation}

Here $E$ is the particle energy in GeV, and
 the symbol $\oplus$ implies that
the stochastic and constant terms in the resolution
are combined in quadrature.

The following values for the stochastic term and the constant
term are extracted from the data:
\begin{equation}
        \sigma_{E}^{pions~mip~in~ECAL}/E = (93.8 \pm 0.9) \%/\sqrt{E} \oplus
                                          (4.4 \pm 0.1)\%
\label{eqn_3}
\end{equation}
for "mip-in-ECAL" pions, and
\begin{equation}
        \sigma_{E}^{all~pions}/E = (82.6 \pm 0.6) \%/\sqrt{E} \oplus
                                          (4.5 \pm 0.1)\%
\end{equation}
for the "full pion sample".

\section{Performance of the HCAL calorimeter with a PbWO$_4$ crystal ECAL}

In this section  the linearity of response and energy resolution
of the HCAL calorimeter with a PbWO$_4$ crystal matrix ECAL are discussed.
The performance of the combined PbWO$_4$ ECAL+HCAL calorimetric  system
was first investigated during 1995 in the H4 beamline. 

\begin{figure}
\begin{center}
\epsfxsize=6in
\mbox{\epsffile{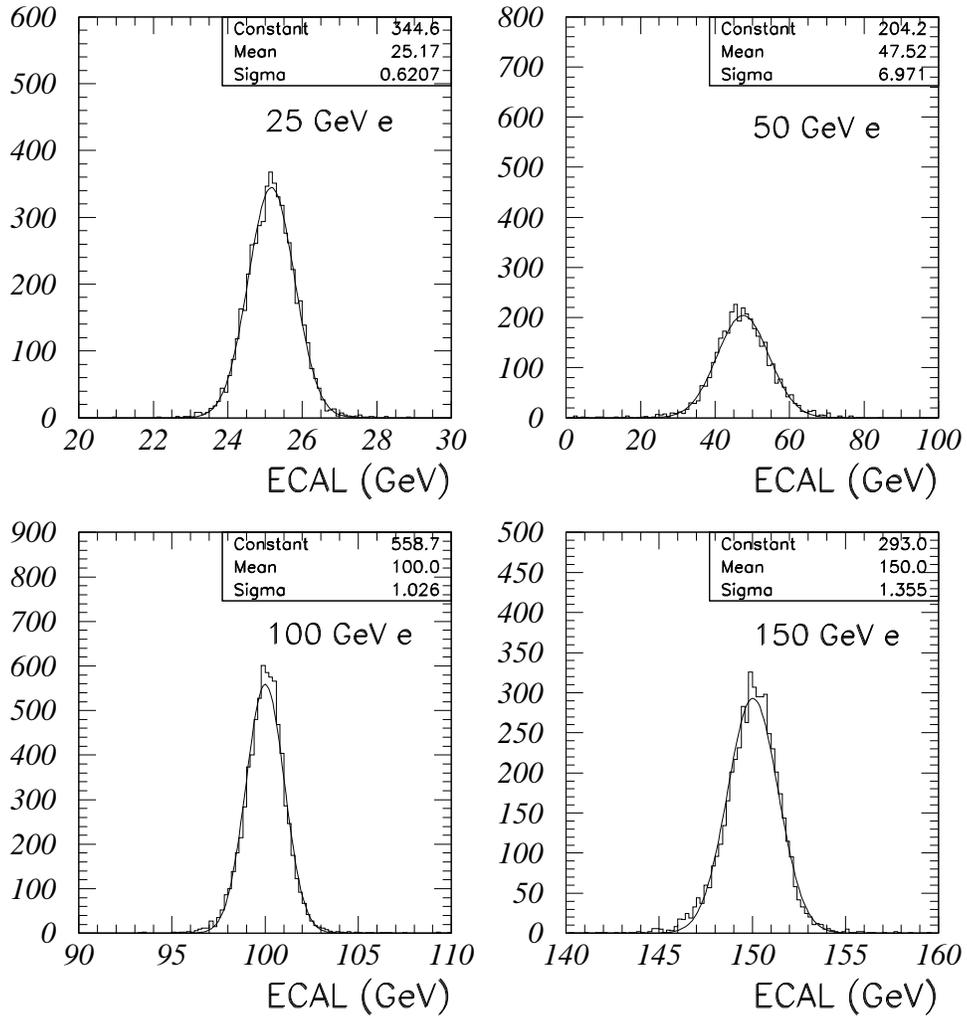}}
\caption{H4(1995) data: response of the PbWO$_4$ crystal ECAL
calorimeter to 25, 50, 100 and 150 GeV/c electrons.}
\label{h4_1995_ele}
\end{center}
\end{figure}

The calibration of the ECAL crystals uses 50 GeV/c electrons
directed into the center of each crystal.
Figure~\ref{h4_1995_ele} shows the energy response of ECAL
for 25, 50, 100, and 150 GeV/c electrons using the 7 $\times$ 7 ECAL
matrix sum.
The energy resolutions of the crystal
ECAL at these energies are found to be
 0.6 GeV, 0.7 GeV, 1.0 GeV, and 1.4 GeV, corresponding to
 fractional energy resolutions of
2.4\%, 1.4\%, 1.0\%, and 0.9\%, respectively.
During these tests, the electronic noise of the 
7$\times$7 crystal matrix energy sum had a rms width equivalent to 440 MeV.
Much better performance of the crystal ECAL calorimeter
has been attained~\cite{ecal_nim} in other tests which were dedicated to
special ECAL studies.

The energy response 
of the combined crystal ECAL+HCAL calorimeter
system is calculated as follows:
\begin{equation}
	E_{TOT} =  \sum_{i}^{ECAL}  \frac{ADC_i}{e_i} /Escale +
                   \sum_{j}^{HCAL}  \frac{ADC_j*w_j}{\mu_j} /Hscale
\end{equation}

Here, $Escale$ and $Hscale$ are constants which define
the absolute energy scales of ECAL and HCAL,
$ADC_i$ are the ADC values of each
readout layer (tower) of the HCAL (ECAL) calorimeter.
The relative calibration
constants of each tower of ECAL,
 $e_i$, correspond to the average 50 GeV/c electron response
of a tower expressed in units of ADC counts.
The coefficients $w_j$ and $\mu_j$ are defined as before (Eqn. 1).

\begin{figure}
\begin{center}
\epsfxsize=6in
\mbox{\epsffile{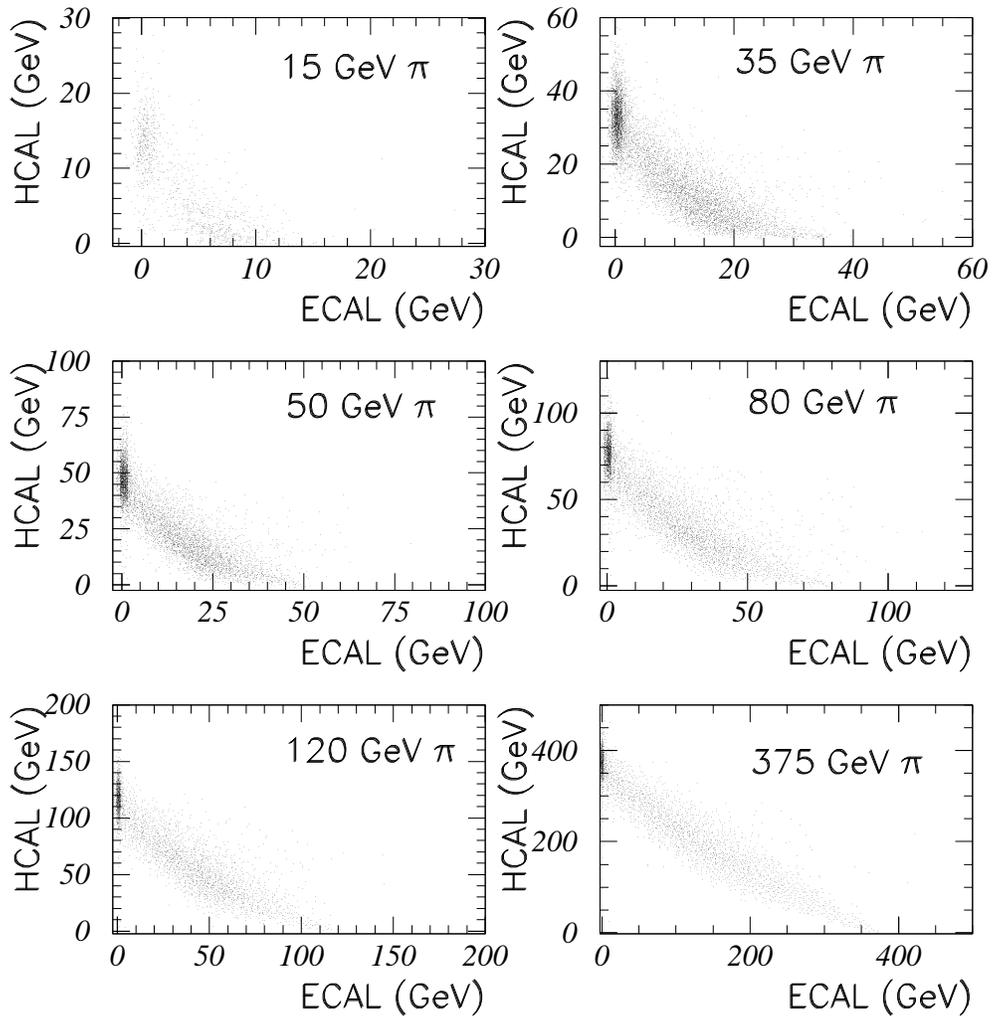}}
\caption{H4(1995) data: scatter plots of the energy in HCAL versus the
energy in the PbWO$_4$ crystal ECAL calorimeter
for pions.}
\label{h4_1995_ecal_vs_hcal}
\end{center}
\end{figure}

\begin{figure}
\begin{center}
\epsfxsize=6in
\mbox{\epsffile{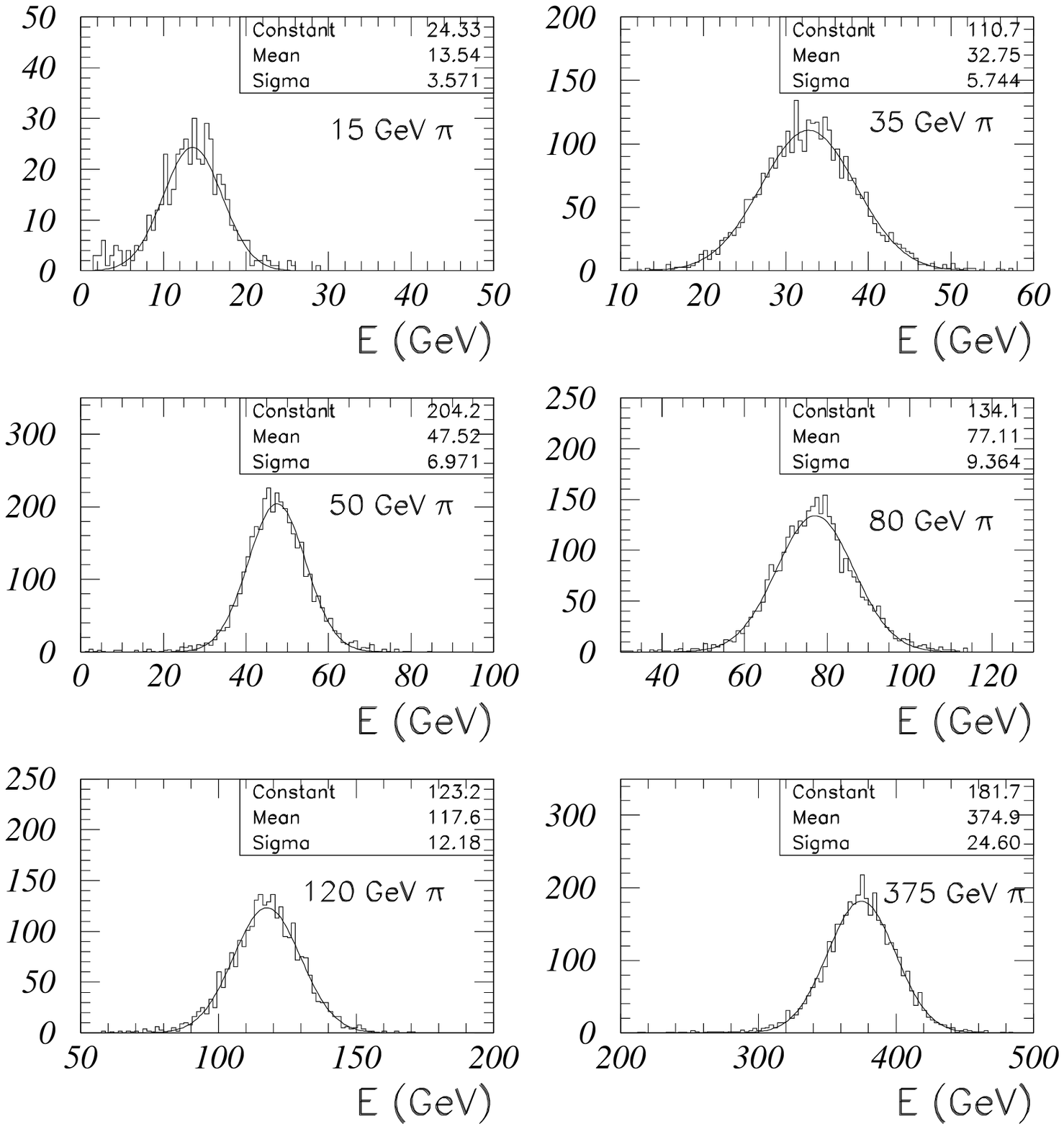}}
\caption{H4(1995) data: the energy response of HCAL for
"mip-in-ECAL" pions.
The minimum ionizing energy deposition in ECAL
has been added to the energy sum.}
\label{h4_1995_etot_mip}
\end{center}
\end{figure}

\begin{figure}
\begin{center}
\epsfxsize=6in
\mbox{\epsffile{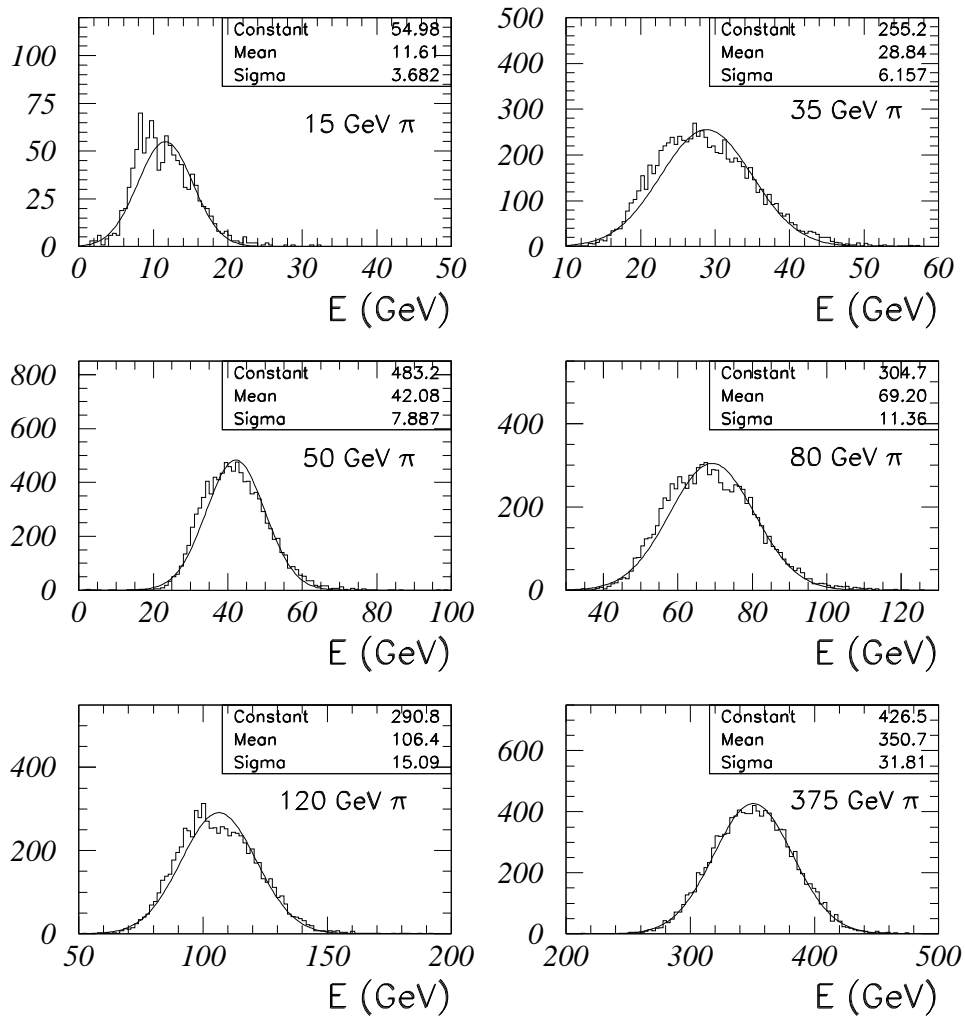}}
\caption{H4(1995) data: the energy response of the combined
crystal ECAL+HCAL
calorimeter to "full pion sample".
}
\label{h4_1995_etot_all}
\end{center}
\end{figure}

Figure~\ref{h4_1995_ecal_vs_hcal} shows
a scatter plot of the energy deposition in 
HCAL versus ECAL  taken during the H4(1995) test beam run
with pions with momenta ranging between 15 and 375 GeV/c.
Unlike in the case with the Pb-scintillator
ECAL, the data points are not centered on a
straight line corresponding to $E_{ECAL}+E_{HCAL}$ = $p_{beam}$.
The energy response of the HCAL calorimeter
to "mip-in-ECAL" pions is shown in
Figure~\ref{h4_1995_etot_mip}.
The response of the combined crystal ECAL+HCAL
calorimeter to the "full pion sample" is shown in
Figure~\ref{h4_1995_etot_all}.
The distributions for the combined crystal ECAL+HCAL system
deviate from a Gaussian shape, especially at low energies.

\begin{figure}
\begin{center}
\epsfxsize=4.5in
\mbox{\epsffile{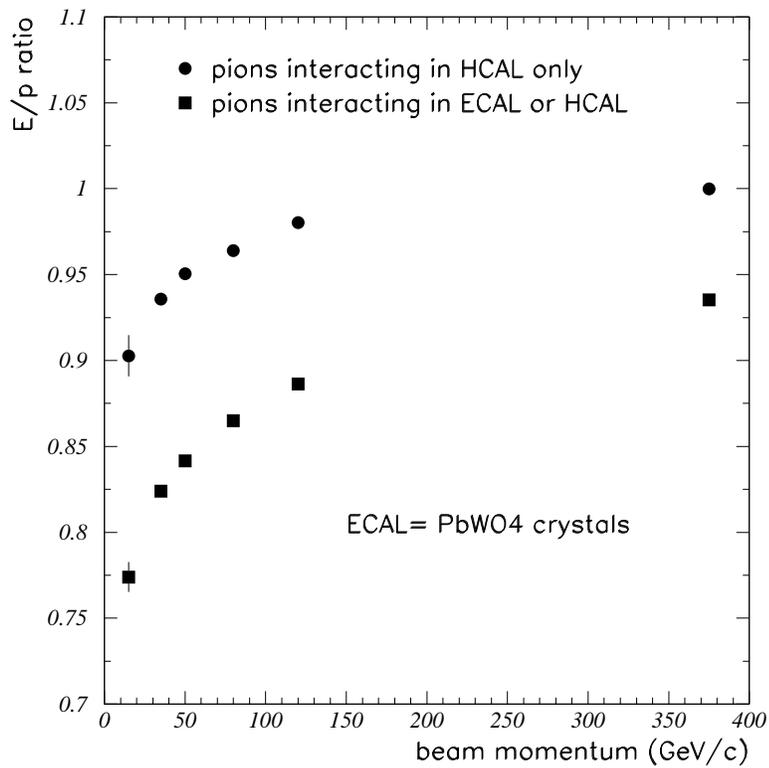}}
\caption{H4(1995) data: linearity of the energy response for
"mip-in-ECAL" and "full pion sample".
The statistical errors for some of the data points are smaller than the symbol 
size.}
\label{h4_1995_lin}
\end{center}
\end{figure}

As demonstrated earlier, the H2(1995) data with a Pb-scintillator ECAL
show good linearity for the combined ECAL+HCAL calorimeter system. 
However, as can be seen in Figure~\ref{h4_1995_lin},
the H4(1995) data with a PbWO$_4$ crystal ECAL is 
quite non-linear. In addition,
the average response of the combined crystal ECAL+HCAL 
calorimeter to "full pion sample" is approximately 10\% lower
than the response to pions which interact in HCAL only.

\begin{figure}
\begin{center}
\epsfxsize=4.5in
\mbox{\epsffile{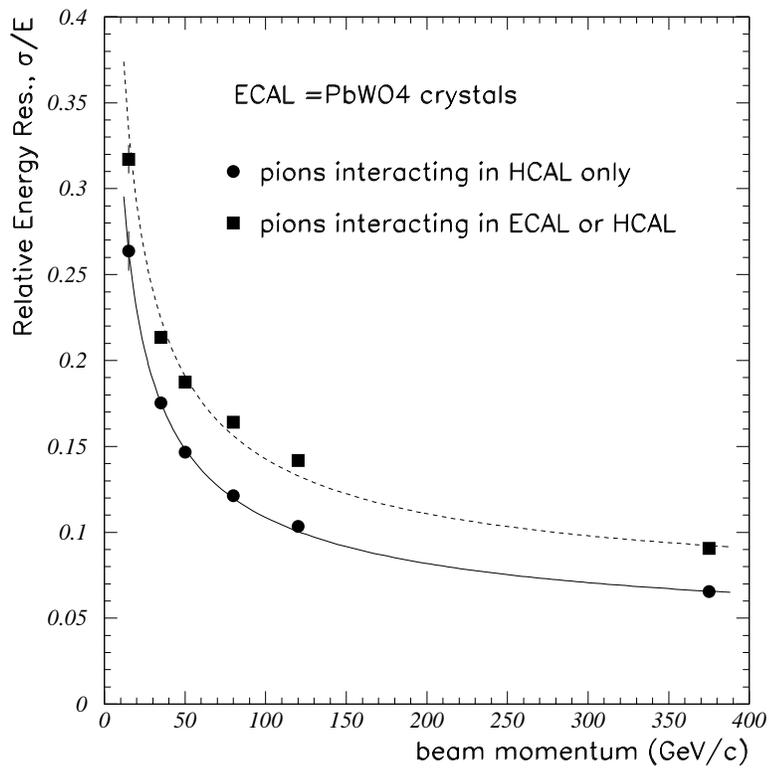}}
\caption{H4(1995) data: fractional energy resolutions for
"mip-in-ECAL" and "full pion sample".
The statistical errors for some of the data points are smaller than 
the symbol size.}
\label{h4_1995_res}
\end{center}
\end{figure}

Figure~\ref{h4_1995_res} shows the fractional energy  resolutions
of the calorimeter for  the "mip-in-ECAL" pions
for  the "full pion sample".
The energy resolutions of the combined crystal ECAL+HCAL 
calorimeter system are significantly degraded, especially at
high energies. These results are attributed to the large
difference between the response of the crystal ECAL to electrons
and hadrons, i.e. the large $e/h$ (the  ratio of the response to the
electromagnetic and  hadronic components of the shower)
of the PbWO$_4$ crystals. This large $e/h$
results in an overall response
of the calorimeter which is very sensitive to fluctuations in the
initial electromagnetic component of hadronic showers.
 
A fit to data for the fractional energy resolution
yields the following values for the stochastic term and constant
term parameters:
\begin{equation}
        \sigma_{E}^{pions~mip~in~ECAL}/E = (101 \pm 0.1) \%/\sqrt{E} \oplus
                                          (4.0 \pm 0.1)\%
\label{eqn_6}
\end{equation}
for the case of "mip-in-ECAL" pions , and
\begin{equation}
        \sigma_{E}^{all~pions}/E = (127 \pm 0.7) \%/\sqrt{E} \oplus
                                          (6.5 \pm 0.1)\%
\end{equation}
for the case of the "full pion sample".
Note that the first fit ("mip-in-ECAL" data , eqn.~\ref{eqn_6})
is consistent with the H2(1995) "mip-in-ECAL" data (eqn.~\ref{eqn_3}).

\section{Response of HCAL to electrons and the $e/h$ ratio for HCAL}

The linearity and energy resolution of hadron calorimeter 
depends~\cite{wigmans} on the intrinsic
$e/h$. The average  fraction of the energy in the electromagnetic
component of hadron induced showers, F($\pi^0$),
 increases as a function
of incident energy. For 
non-compensating calorimeters ( $e/h$ $\neq$1), this implies a non-linear
hadron energy response. 
The event-by-event fluctuations
in F($\pi^0$) contribute to variations in the reconstructed
shower energy and 
dominate the energy resolution for
hadrons at high energies.

In order to study the relative response of HCAL to 
electron and pions, data were taken with ECAL removed
from the beamline. In order to reject electrons present in the hadron
tunes from the
sample, a minimum ionizing signal was required in the
first three sampling layers of HCAL.

\begin{figure}
\begin{center}
\epsfxsize=4.5in
\mbox{\epsffile{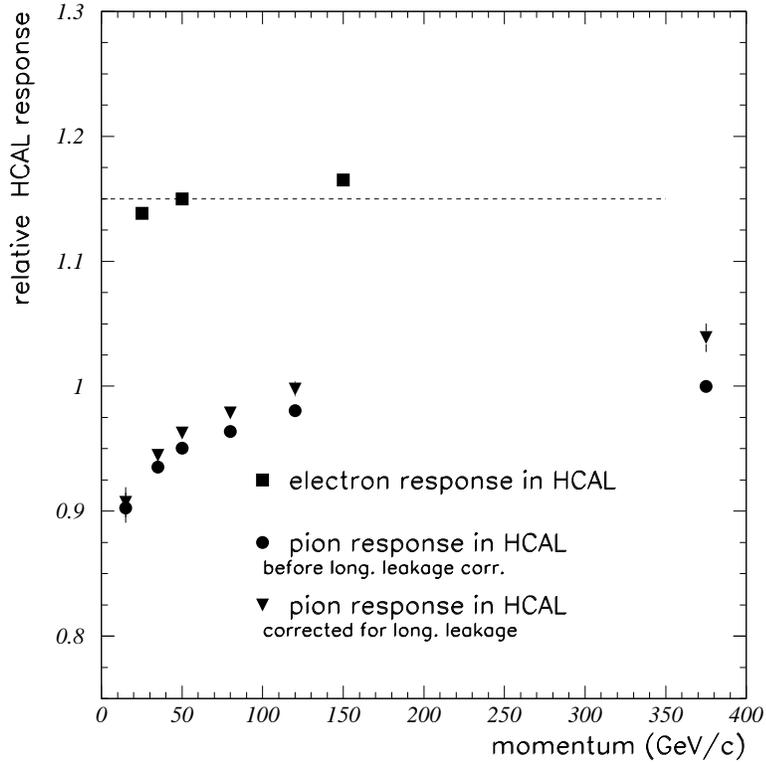}}
\caption{H4(1995) data: the 
response of HCAL to electrons. The ECAL crystals are moved out
of the beamline for these data.
The figure also shows the response
of HCAL to pions before and after
a correction for downstream longitudinal energy leakage is applied.
A 30\% systematic uncertainty is assumed on the longitudinal
leakage correction. The dotted line represents the ratio of the 
average electron response to the pion response at 375 GeV/c.
}
\label{eh-cms1}
\end{center}
\end{figure}

The studies of the energy response of HCAL to pions
indicate that the HCAL calorimeter is somewhat non-linear.
There is $\approx$ 9\% increase in the relative response 
of  HCAL to pions between 15 and 375 GeV/c, as shown in
Figure~\ref{eh-cms1}. 
The total interaction length~\footnote{Taking into
account the requirement that energy deposition in first three HCAL counters
is consistent with minimum ionizing particles.}
of the H4(1995) HCAL module
is approximately
 8 $\lambda_{INT}$. At very high energies, an  8 $\lambda_{INT}$
calorimeter is not expected to fully contain the hadronic shower.
The correction for longitudinal leakage
for an 8 $\lambda_{INT}$ calorimeter is estimated~\cite{nutev_lucy} to be
1.5\% at 120 GeV/c and 3.5\% at 375 GeV/c, using CCFR
~\cite{ccfr_nim} and NuTeV~\cite{nutev_nim}
test beam data. Therefore, the intrinsic non-linearity
of HCAL is somewhat larger if the corrections for leakage at high
energies are applied.

The response of the H4(1995) HCAL module to
electrons is linear at a level of 2\% and is approximately 15\% higher
than the average response to pions as shown in Figure~\ref{eh-cms1}.
The absolute energy scale of HCAL is set (as before)
by the 375 GeV/c pion data point.

The electron and
pion data  can be used to extract $e/h$ by applying  the following expressions:

\begin{equation}
\pi = F(\pi^0)\times e + (1-F(\pi^0))\times h
\end{equation}

\begin{equation}
\frac{\pi}{e} = \frac{1+(\frac{e}{h} -1) \times F(\pi^0)}
                      {\frac{e}{h}}
\label{pie_eqn}
\end{equation}

Here $e$ is the response of the calorimeter to the electromagnetic
components of hadron showers 
and $h$ is the response to the
hadronic component of pion showers.
$\pi$ is the response
to pions and  F($\pi^0$) is the fraction of the energy in
electromagnetic component of pion
showers. 

The extracted value of $e/h$ depends on the assumed
parameterization  for F($\pi^0$)
as a function of energy.
The Wigmans~\cite{wigmans} formula for F($\pi^0$) (eqn.~\ref{wig_eqn}) 
increases with energy
($\approx$ log(E)) and becomes non-physical at very high energies.
The Groom parameterization (eqn.~\ref{groom_eqn}) is
extracted from a CALOR Monte Carlo simulation~\cite{groom} and
uses a power law dependence.

\begin{equation}
F_W(\pi^0) = 0.11 \times ln(E)
\label{wig_eqn}
\end{equation}

\begin{equation}
F_G(\pi^0) = 1 - (E/0.96)^{0.816-1}
\label{groom_eqn}
\end{equation}

\begin{figure}
\begin{center}
\epsfxsize=4.5in
\mbox{\epsffile{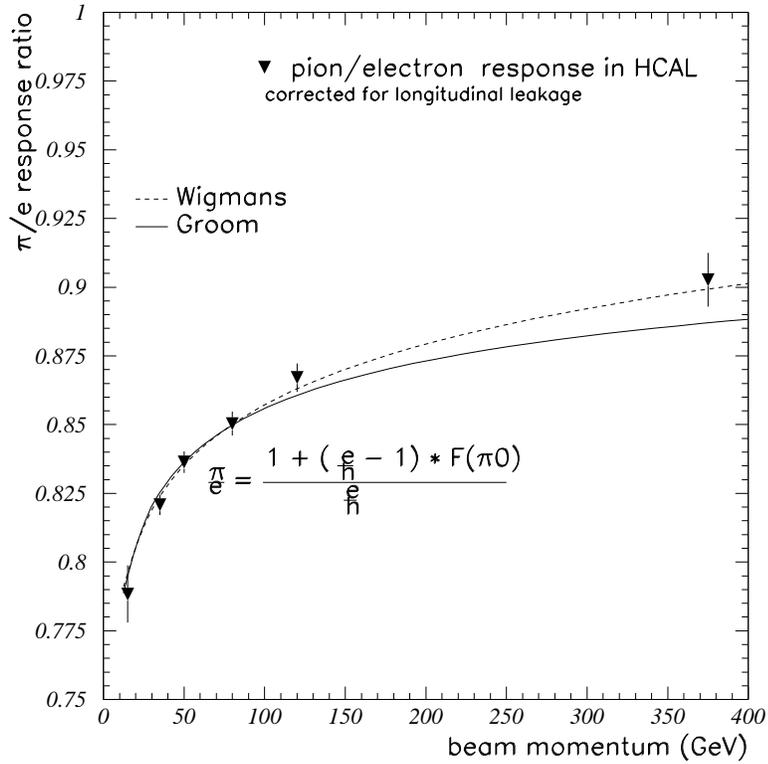}}
\caption{H4(1995) data: the pion/electron response ratio for
HCAL (only) as a function of beam momentum.
The pion response of HCAL has been corrected for longitudinal
leakage. 
We  assume a linear electron response for HCAL (only), 
and use  E(ele)/E($\pi$)=1.15$\pm$0.01 at 375 GeV/c. 
}
\label{eh-cms2}
\end{center}
\end{figure}

Using the above two parameterizations, we extract values of $e/h$ 
of 1.41 $\pm$ 0.01 (Wigmans) and 1.51 $\pm$ 0.01 (Groom). 
The best fit to our data is shown in  Figure~\ref{eh-cms2}. 

The above results can be compared with similar analyses for other
iron-scintillator sampling calorimeters.
Extraction of $e/h$ for an iron-scintillator  sampling
calorimeter has also
been done~\cite{scifi97} by CDF.
The CDF endcap hadron calorimeter consists of 22 layers of 5~cm iron
absorber
plates and 6~mm scintillators.
The CDF results yield 
$e/h$ 1.34 $\pm$ 0.01 (Wigmans) and 1.42 $\pm$ 0.015 (Groom).
ATLAS~\cite{atlas_nim} measures $e/h$= 1.34 $\pm$
0.03
assuming the Wigmans parameterization for $F({\pi^0})$.
The ATLAS hadron calorimeter uses
iron as an absorber with
the tiles are oriented in the radial
direction.

The NuTeV  Collaboration 
reports a $e/h$ ratio for 10~cm iron sampling
calorimeter~\cite{nutev_nim} of 1.08 $\pm$ 0.01 using the
Groom parameterization. Note that $e/h$ is sensitive to the sampling
fraction and other geometrical effects. The CCFR/NuTeV calorimeter
uses 2.5 cm thick liquid scintillator counters, which are clad with
acrylic and water bags. The cladding and the thick scintillators in 
CCFR/NuTeV greatly reduce the response to electrons relative to the response
to pions. Also, the water and thick scintillators
tend to increases the rate of conversion of low energy neutrons to protons
in the active medium.

\section{Optimization of the design of HCAL}

The conceptual design of the barrel HCAL was a 5.3~$\lambda_{INT}$ 
thick calorimeter. The inner half of the calorimeter had 3~cm
sampling (first readout segment, H1) while the second readout 
segment consisted of
the outer half of the calorimeter with 6~cm sampling. 

	For the H2(1996) tests, the prototype HCAL module was segmented 
into 27 readout layers. Using these data we simulated various sampling 
configurations and studied the performance of HCAL as a function of 
total interaction length and sampling frequency. These data allowed us to 
optimize the calorimeter response to pions and jets, 
while taking into account the existing design 
(e.g. geometrical and readout) constraints imposed on HCAL within the 
CMS detector.

Important design choices that were made based on these data are: 
\begin{itemize}
\item the absorber sampling thickness; 
\item the depth  of the HCAL inside the magnet (in interaction
lengths); 
\item the longitudinal segmentation of the inner HCAL into H1 and H2
readout segments; 
\item possibility of adding scintillator layers outside of the magnet to 
create an outer calorimeter.
\end{itemize}

\subsection{Choice of the longitudinal segmentation for HCAL}

Studies performed
   prior to
   the H2(1996) beam tests resulted in  no compelling argument to set the
   optimal partition between the two readout segments of inner HCAL.
   However,
   the 1996 test beam data indicate that the proper partition is 
   that which is most useful in correcting for large $e/h$ response
   of the ECAL crystals. 
To compensate for a large $e/h$ of the combined ECAL + HCAL
system, 
we chose to make a novel longitudinal segmentation, 
with the first depth segment, H1,  reading out only the first scintillator
layer. 
The remaining scintillator layers inside the magnet are combined 
into the second readout, H2.
   The reason for  this choice is the following. 
   A large H1 signal indicates that a significant amount of hadronic energy 
   has been deposited  in ECAL and is  underestimated
   because of the large $e/h$ of the crystals.
   The information from H1 can be used to
   make an effective correction for the hadronic non-linearity of
   the ECAL crystals.

\subsection{Response of the combined crystal ECAL+HCAL calorimeters}

\begin{figure}
\begin{center}
\epsfxsize=4.5in
\mbox{\epsffile{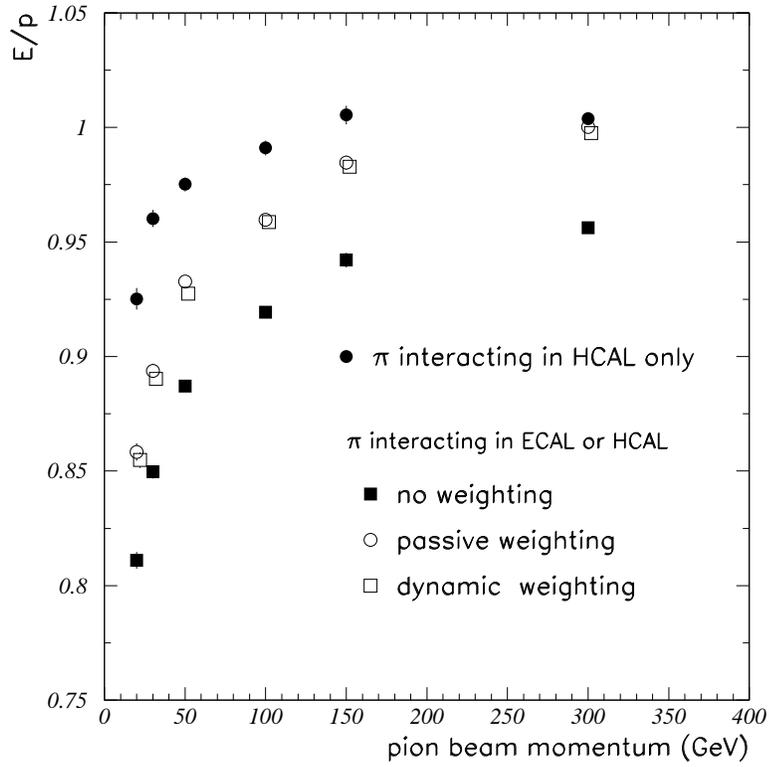}}
\caption{H2(1996), B = 0 Tesla data: the
linearity of the response of HCAL to 
"mip-in-ECAL" pions
and of the combined PbWO${_4}$ crystal ECAL + HCAL system to 
"full pion sample"
with $\alpha$=1.4 and without the weighting of H1.
The statistical errors are smaller than the size of the symbols.
}
\label{lin}
\end{center}
\end{figure}

\begin{figure}
\begin{center}
\epsfxsize=4.5in
\mbox{\epsffile{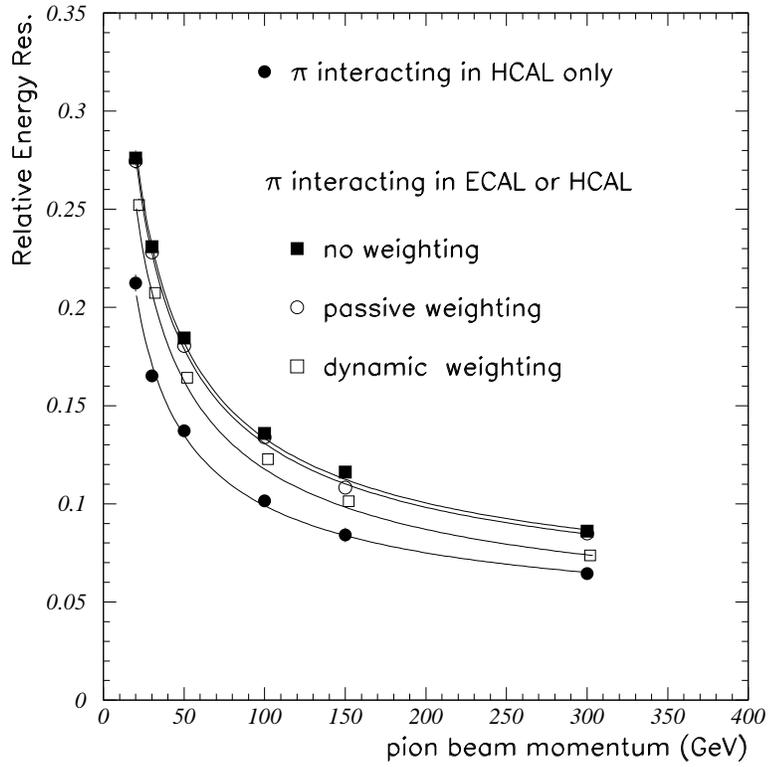}}
\caption{H2(1996), B = 0 Tesla data: the
fractional pion energy resolution  of HCAL to
"mip-in-ECAL" pions and
of the combined response of the PbWO$_4$ crystal ECAL+HCAL system to
"full pion sample"
with $\alpha$=1.4 and without the weighting of H1. The statistical errors 
are smaller than the size of the symbols.
}
\label{res}
\end{center}
\end{figure}

Figure~\ref{lin} and~\ref{res} show 
the energy response and fractional energy resolutions
of the combined PbWO$_4$ crystal ECAL+HCAL calorimeters to pions. 
The inner HCAL(5.3~$\lambda_{INT}$)
consists of two independent readouts: H1 (following
a 2~cm Cu plate) and H2 (thirteen 6~cm Cu samplings).
A single readout outer calorimeter (HO) consists of three samplings: 
(1) the first sample is immediately after the magnetic
coil (which is mimicked by 18~cm of Cu in the test
beam ),(2) a second 22~cm Cu sampling layer,
and (3) a third 16~cm Cu sampling layer.
The combined coil and HO samplings correspond
to a total of 2.5$ \lambda_{INT}$. 

We have investigated two possible approaches to correct for the degradation
of the performance of the combined crystal ECAL+HCAL calorimeters.
   Both of these methods make use of the segmented readout of  
   HCAL.
   The first approach, called
   passive weighting, 
increases the weight ($\alpha$) of the first (H1) HCAL 
   readout segment, where $\alpha$ is an energy independent constant.
   Note that $\alpha$ =1 corresponds to using the
   standard "absorber Simpson's weighting"
   used elsewhere in the calorimeter.

\begin{equation}
    E_{TOT} = E_{ECAL} + \alpha \times E_{H1} + E_{H2} + E_{HO}
\end{equation}

\begin{figure}
\begin{center}
\epsfxsize=4.5in
\mbox{\epsffile{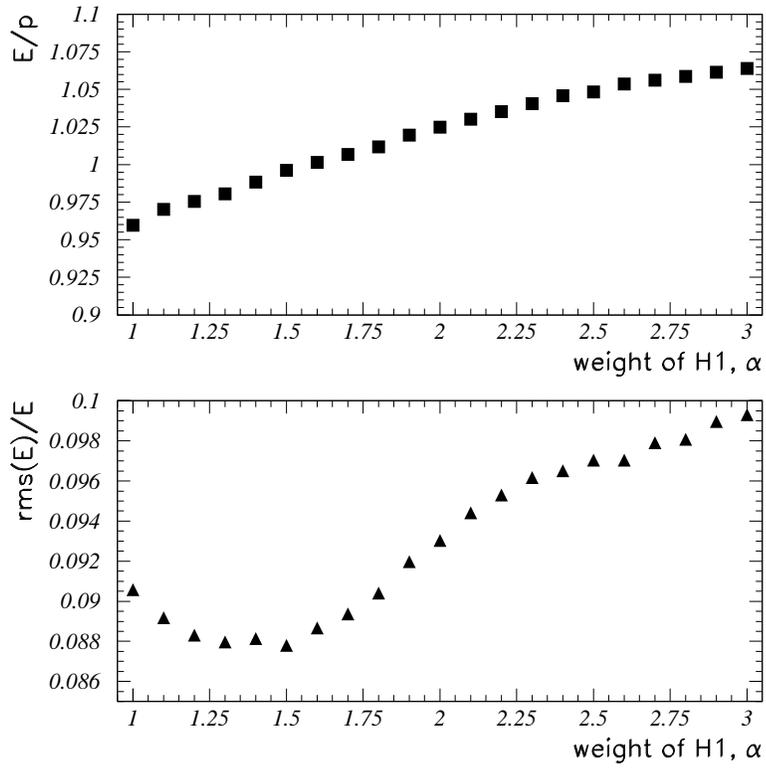}}
\caption{H2(1996) data, B = 0 Tesla data:
the dependence of the $E/p$ and the
fractional rms energy resolution for 300 GeV/c pions as a function of
parameter ${\alpha}$, the weight assigned to
H1 (the first readout compartment of HCAL).
}
\label{tb300b0n4-a}
\end{center}
\end{figure}

   The overall linearity and 
   fractional energy resolution of the combined
   crystal ECAL+HCAL system for 300 GeV/c pions
   as a function of the parameter $\alpha$ are shown
   in  Figure~\ref{tb300b0n4-a}. 
   Both the linearity and
   energy resolution of the combined crystal
   ECAL+HCAL system are improved for the value of $\alpha$ = 1.4.

   The second approach, called dynamic
   weighting is an event-by-event correction. It depends
    on the fraction of the energy
   deposited in the first readout segment of HCAL immediately 
   downstream of the crystal ECAL. The dynamic weighting
   effectively allows one to have an energy dependent
   correction (for single particles) for the low response to pions
   which interact in ECAL. 

   \begin{equation}
    E_{TOT} = (1+2\times f(H1))\times E_{ECAL} + E_{H1} + E_{H2} + E_{HO},
   \end{equation}
   \begin{equation}
   f(H1) = E(H1)/(E(H1)+E(H2)+E(HO)),~f(H1) \leq 0.1
   \end{equation}

   With either the passive or dynamic weighting, 
   the nonlinearity and resolutions
   for pions with energies between 30 and 300 GeV/c are improved.
   With the passive weighing, the fractional energy resolution of 
   of the combined ECAL+HCAL calorimetric system can be described by the
   function $\sigma_{E}/E$ = 122\%/$\sqrt{E}$ $\oplus$ 5\%.
   Note that while the passive weighting
   can be applied to single particles and  jets, 
   the dynamic weighting may introduce
   high energy tails in the case of particle jets.

   Techniques to overcome the problem of the different non-compensation
   properties (i.e. $e/h$)
   of the electromagnetic and hadronic compartments
   for combined calorimetry systems have been 
   studied~\cite{atlas}
   by the  ATLAS calorimetry group. The algorithms proposed
   by the ATLAS group depend on the energy fraction deposited 
   in the electromagnetic compartment and therefore can be only
   applied to single particle energy reconstruction.

\subsection{Studies of the total HCAL absorber depth}

\begin{figure}
\begin{center}
\epsfxsize=4.5in
\mbox{\epsffile{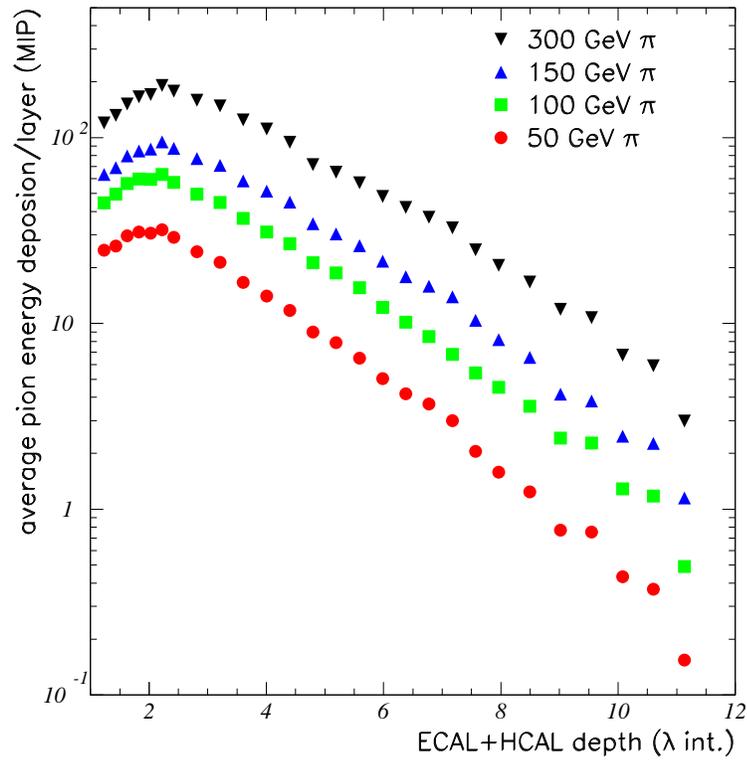}}
\caption{H2(1996), B = 0 Tesla data: average hadron shower profiles for
   50, 100, 150 and 300 GeV/c pions
   as a function of calorimeter absorber depth. The total $\lambda_{INT}$
   includes the contribution of ECAL.}
\label{r-profiles}
\end{center}
\end{figure}

We use the 27 sampling layers of the H2(1996)
test beam module to study the calorimeter
performance as a function of total depth. 
     Figure~\ref{r-profiles} shows the average pion shower profiles as a
function of
total absorber depth for 50, 100, 150, and 300 GeV/c pions. As shown in
the figure, the average pion shower profiles extend significantly
beyond 7~$\lambda_{INT}$, in particular at high pion energies.
Fluctuations in
the leakage for high energy pions also become large.

     To collect this leakage energy, the barrel HCAL has been augmented
with an additional outer calorimeter (HO) consisting of a single layer
of scintillators beyond the solenoid magnet (the solenoid thickness is
1.4~$\lambda_{INT}$).  For pseudorapidity $\eta$ less than 0.4, a layer of 
iron (thickness = 18 cm = 1.1~$\lambda_{INT}$) has been instrumented with
an
additional scintillator layer.  For either case (one or two layers of
scintillator), the scintillators cover the same solid angle as the
interior calorimeter and are read out as a single depth segment. The
combined solenoid + iron of the HO corresponds to an additional 
2.5~$\lambda_{INT}$.

\begin{figure}
\begin{center}
\epsfxsize=4.5in
\mbox{\epsffile{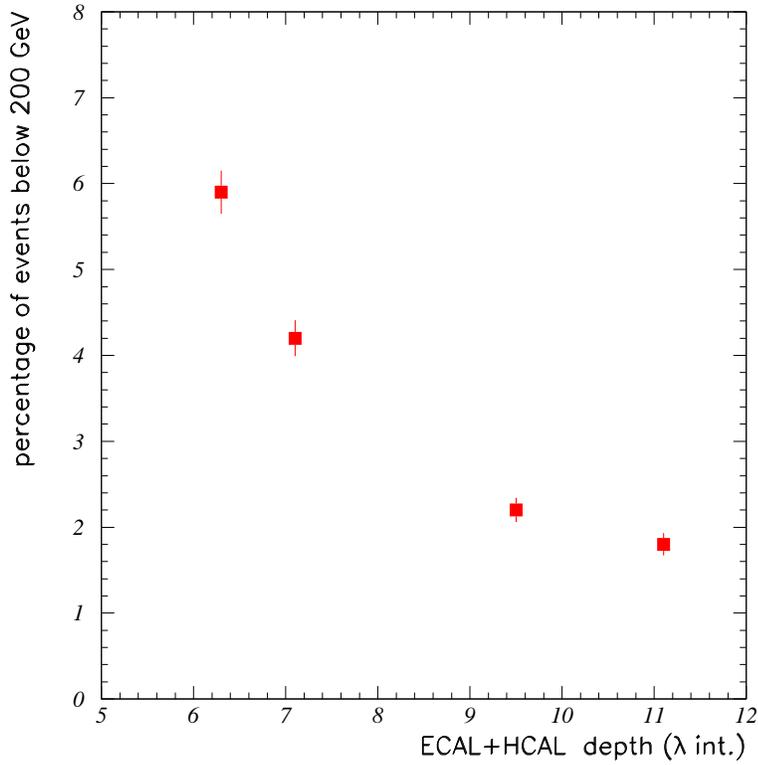}}
\caption{ H2(1996) data, B = 3 Tesla: the
fraction of 300 GeV/c pions
   with  reconstructed energy less than 200 GeV (approximately
   3 $\sigma$ below the mean, or 100 GeV of missing energy) versus
   total absorber depth. The total $\lambda_{INT}$
   includes the contribution of ECAL.}
\label{n3-leakage}
\end{center}
\end{figure}

     The fraction of 300 GeV/c pions with reconstructed energy less than
200 GeV (approximately 3~$\sigma$ below the mean, or 100 GeV of missing
energy) is shown in Figure~\ref{n3-leakage}. The four points correspond
to: 5.9~$\lambda_{INT}$,
the HCAL alone; 7.0~$\lambda_{INT}$, HCAL + ECAL; 9.5~$\lambda_{INT}$,
HCAL
+ ECAL + HO; and 11~$\lambda_{INT}$ which is the total thickness of the 27
layer test beam module. We see that for the CMS barrel design, 
at 9.5~$\lambda_{INT}$, less than 2\% of
the pions are catastrophically mismeasured. 
To increase the interaction length of the calorimeter to
9.5~$\lambda_{INT}$ 
we added two additional
absorber plates inside the magnet with respect to the conceptual design.
This was achieved by 
reducing  the inner radius of barrel HCAL.

\subsection{Optimization of Absorber Sampling Thickness}

\begin{figure}
\begin{center}
\epsfxsize=4.5in
\mbox{\epsffile{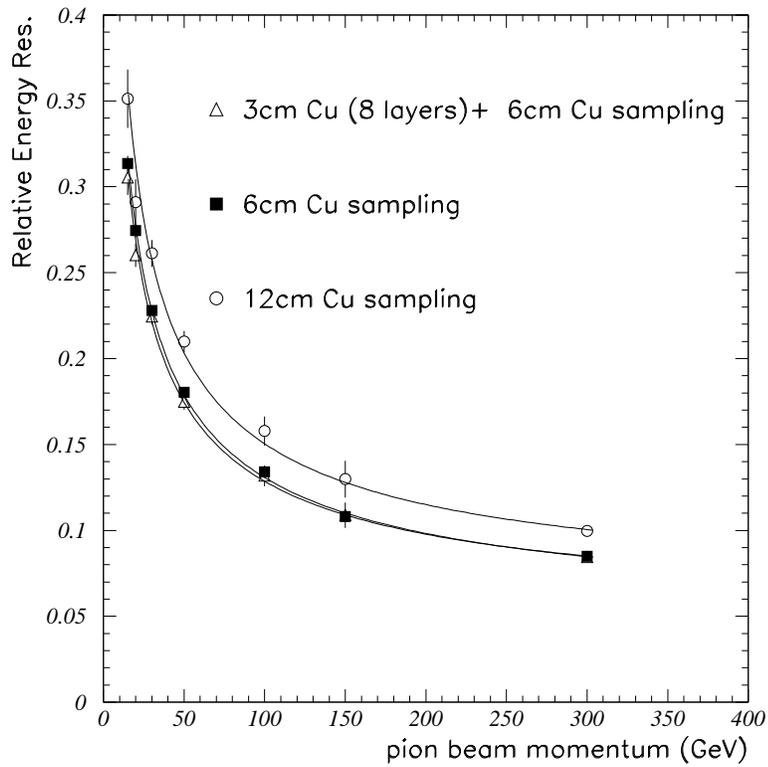}}
\caption{H2(1996), B = 3 Tesla data:
comparison of the fractional pion energy resolution  of the
combined crystal ECAL+HCAL system,
for various choices of
absorber samplings in the inner HCAL:(a) 3~cm Cu sampling for first 8 layers,
followed by 6~cm sampling, (b) 6~cm uniform sampling, and
(c) 12~cm uniform sampling.
}
\label{nb04_nb06_nb07-97}
\end{center}
\end{figure}

The fractional energy resolution has been investigated
for the following three choices for the inner HCAL absorber samplings:
 a) 3~cm Cu sampling for the first eight layers followed by 6~cm Cu
sampling,
 b) 6~cm Cu uniform sampling, and
 c) 12~cm Cu uniform sampling.
The data, shown in Figure~\ref{nb04_nb06_nb07-97} 
indicate that the energy resolution is not dominated
by the sampling fluctuations in HCAL. 
A factor of two decrease in the sampling
frequency ( 3~cm/6~cm to 6~cm uniform sampling)
for HCAL does not result in a noticeable
degradation of the energy resolution of the
combined detector system. In the case 
of  12~cm Cu uniform sampling, the degradation in energy resolution is
noticeable, but does  not scale with $\sqrt{t}$, where $t$ is the
thickness of the absorber plates. 
For the final design, we chose
uniform 5~cm
sampling which does not degrade the energy resolution and puts more
absorber
inside the magnet, relative to the conceptual design.

\subsection{Final Design of CMS Barrel HCAL}

The final design of the CMS barrel HCAL consists of 17
absorber/scintillator samples
with 5 cm absorber thickness. The total thickness of absorber at 90 degrees 
is 5.9~$\lambda_{INT}$.
With the ECAL included, the total thickness inside 
the magnet is 7.0~$\lambda_{INT}$.
Exterior to the magnet, there  is a 2.5~$\lambda_{INT}$ outer
calorimeter. 
Each projective tower has 3 readouts in depth: the very 
first scintillator layer (H1); the remaining 16 scintillator layers 
inside the magnet (H2); and the outer calorimeter (HO).

\section{Monte Carlo simulation of the test beam results}

  Several hadron shower generators are available
  within the GEANT~\cite{ref_geant} framework. These include
  GHEISHA~\cite{gheisha}, GFLUKA (which is an implementation
  of FLUKA~\cite{fluka} within GEANT),
  and GCALOR (which is an implementation
  of CALOR~\cite{calor} within GEANT). Although GHEISHA is native
  to the GEANT program, GFLUKA and GCALOR are imperfect implementations
  of the original FLUKA and CALOR programs. Both of these programs have
  been known to produce somewhat different
  results ~\cite{imperfect} than the original generators.

  GEANT is used in various studies for evaluation of  the calorimeter
design
  for  CMS. In order to verify those simulations and to understand their
  limitations, GCALOR is used to simulate the H2(1996) test beam data. It
  also serves  as a reference for comparison with the other generators. 
  Our aim is first to anchor the Monte Carlo model to the ensemble of
  test beam data. Once it is so constrained, it is assumed that it can be
  used to make small extrapolations to model the final CMS calorimeter
  system. 

  Details of the crystal ECAL and HCAL test beam geometry 
  are implemented in the GCALOR simulation. For ECAL, this 
  includes a  7 x 7 matrix of individual
  crystals surrounded by copper blocks as well as mechanical and cooling
  structures.
  The HCAL geometry includes the layer structure
  of copper plates, scintillator, plastic cover plates, and air gaps 
  on both sides  of each scintillator. In order to take into account all 
  experimental effects, the transverse beam profiles are simulated using
  information from the test beam tracking chamber.
  Electronic noise and photo-statistics effects
  are simulated based on the measured distributions of pedestals,
electron,
  and muon signals in ECAL and HCAL.
   The longitudinal light collection efficiencies in the
  central nine crystals are
  included in the simulation. The energy cut values
  in the GEANT simulation
  are set to the GEANT
  default values, 1 MeV for electrons and 10 MeV for hadrons.

\begin{figure}
\begin{center}
\epsfxsize=4.5in
\mbox{\epsffile{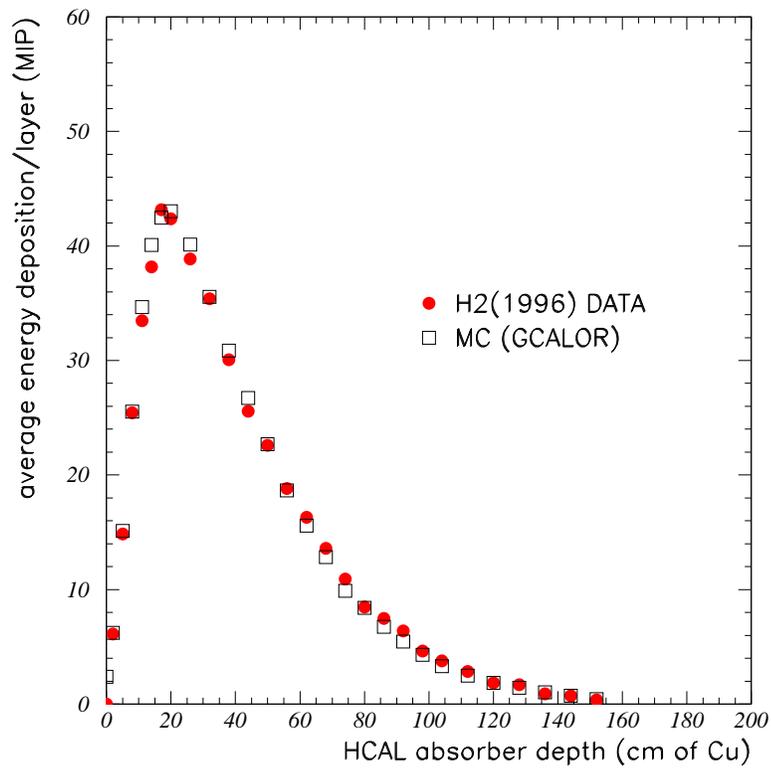}}
\caption{
Comparison of H2(1996), B = 0 Tesla data
with  GEANT-GCALOR simulations
for the average longitudinal shower profile of 
50 GeV/c pions.
The electromagnetic calorimeter 
has been taken out of the beamline.
}
\label{h0-data-geant-prf}
\end{center}
\end{figure}

\begin{figure}
\begin{center}
\epsfxsize=4.5in
\mbox{\epsffile{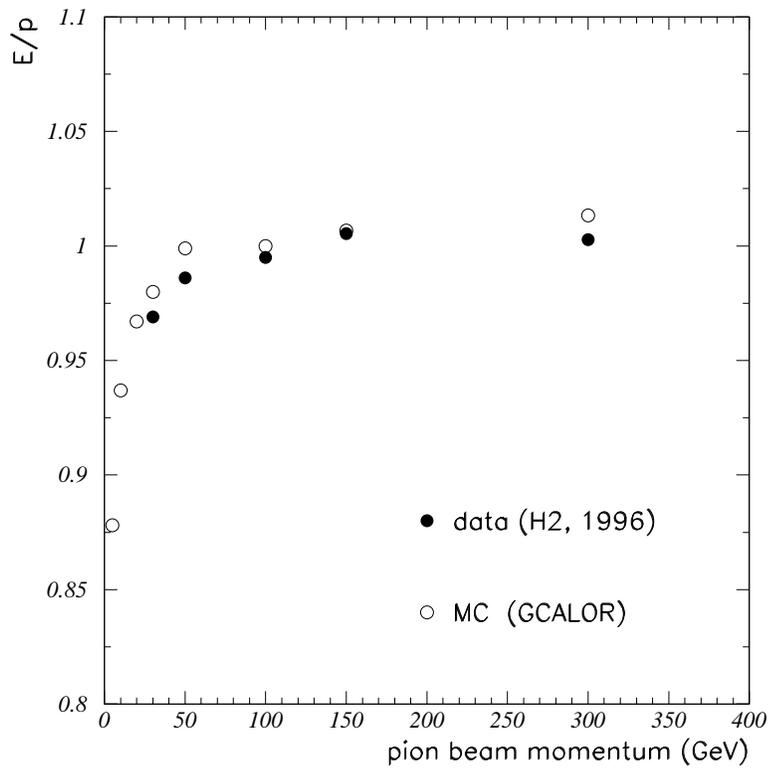}}
\caption{ H2(1996), B = 0 Tesla data: linearity of
the energy response of the hadron calorimeter (HCAL only) 
for pions and comparison with the
GEANT-GCALOR MC simulation.  
The electromagnetic calorimeter 
has been taken out of the beamline.
}
\label{h0-data-geant-lin}
\end{center}
\end{figure}

\begin{figure}
\begin{center}
\epsfxsize=4.5in
\mbox{\epsffile{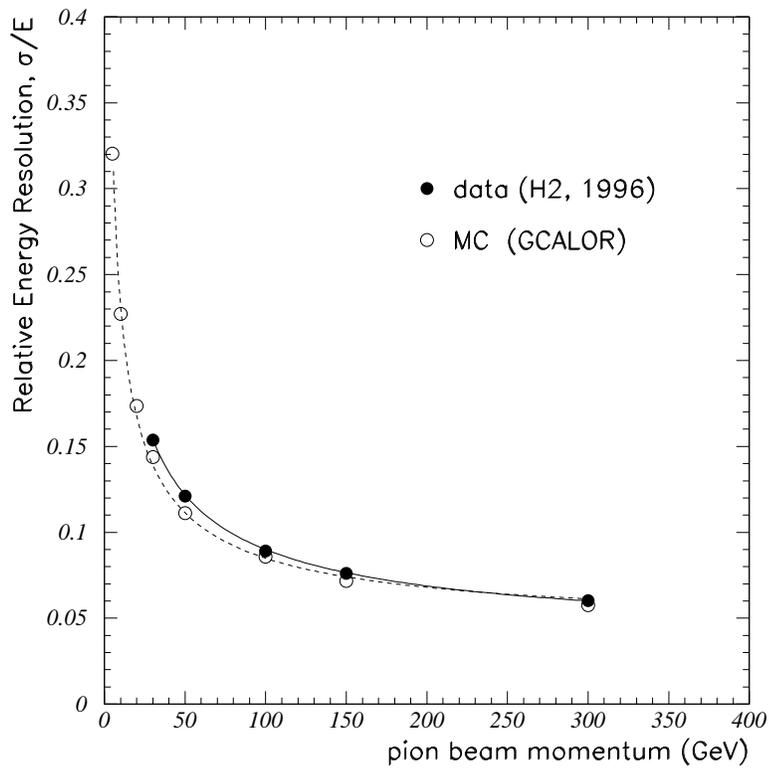}}
\caption{H2(1996), B = 0 Tesla data: the fractional energy resolution
of the  hadron calorimeter (HCAL only) for
pions  and comparison with the GEANT-GCALOR MC simulation. 
The electromagnetic calorimeter has
been taken out of the beamline.}
\label{h0-data-geant-res}
\end{center}
\end{figure}

  In order to compare response of pions interacting in HCAL,
  we have removed ECAL from the beam.
  For this dataset, the test beam data are in good agreement with
  the results of the GCALOR Monte Carlo simulations.
   Good agreement is observed in the longitudinal shower
  profile (Fig.~\ref{h0-data-geant-prf}), the pion response versus
  energy (Fig.\ref{h0-data-geant-lin}), and in the fractional
  pion energy resolution (Fig.\ref{h0-data-geant-res}). 

\begin{figure}
\begin{center}
\epsfxsize=4.5in
\mbox{\epsffile{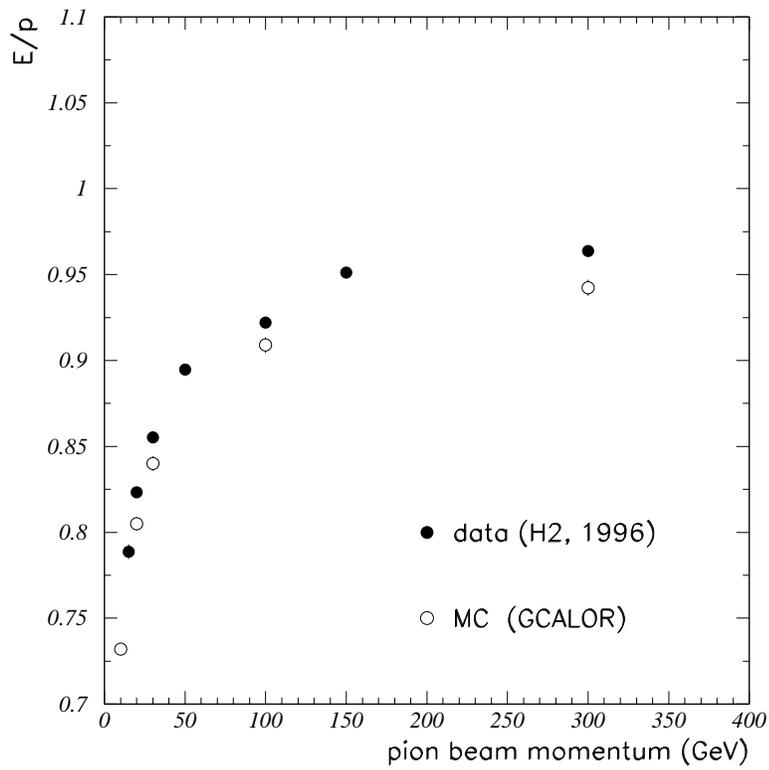}}
\caption{H2(1996), B = 0 Tesla data: the linearity of the energy response
of the PbWO$_4$ 
crystal ECAL+HCAL combined system for the "full pion sample"
and comparison with the GEANT-GCALOR MC simulation.}
\label{n0-data-geant-lin}
\end{center}
\end{figure}

\begin{figure}
\begin{center}
\epsfxsize=4.5in
\mbox{\epsffile{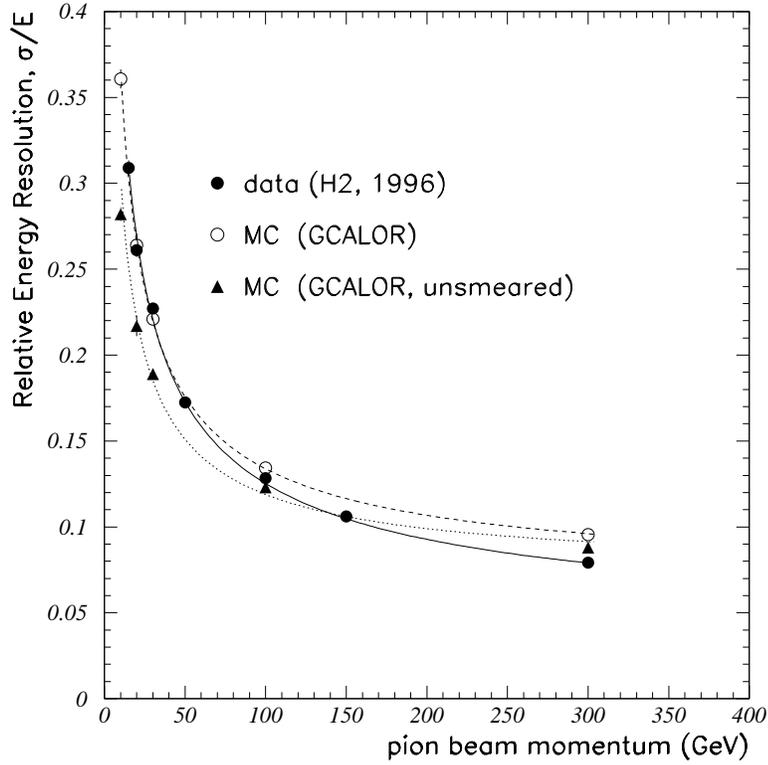}}
\caption{H2(1996), B = 0 Tesla data: the energy resolution  of pions  of
the PbWO$_4$ 
crystal ECAL+HCAL combined system for the "full pion sample"
and comparison with the GEANT-GCALOR 
MC simulation.
The MC simulation includes test beam detector effects
(such as transverse leakage from prototype ECAL and
ECAL electronics noise).
Also shown (triangle symbols) is the MC simulation of 
the energy resolution  of pions in crystal HCAL+ECAL combined system
without these test beam detector effects (which are not expected
to be present in the CMS experiment).
}
\label{n0-data-geant-res}
\end{center}
\end{figure}

  A comparison of the linearity and fractional energy resolutions
  for the "full pion sample" using the combined crystal ECAL+HCAL system
for test beam data
  and the  GEANT simulations are shown
  Figures~\ref{n0-data-geant-lin} and~\ref{n0-data-geant-res}.
  The GEANT simulations, which include all experimental detector effects
  such as electronics noise, are in good
  agreement with the test beam data.
  Also  shown 
  in Figure~\ref{n0-data-geant-res} are the results of a MC simulation
  of the crystal ECAL+HCAL energy resolution
  excluding test beam detector effects,
  such as the ECAL electronic noise, and the 
  energy leakage from the small prototype ECAL.
  As can be seen  in Figures~\ref{n0-data-geant-res}, the detector effects
  present during the H2(1996) test beam data taking significantly 
  degrade the fractional pion energy
  resolution, especially at the low energies.
  These test beam detector effects are not expected to be present
   in the final CMS configuration.
 
\begin{figure}
\begin{center}
\epsfxsize=4.5in
\mbox{\epsffile{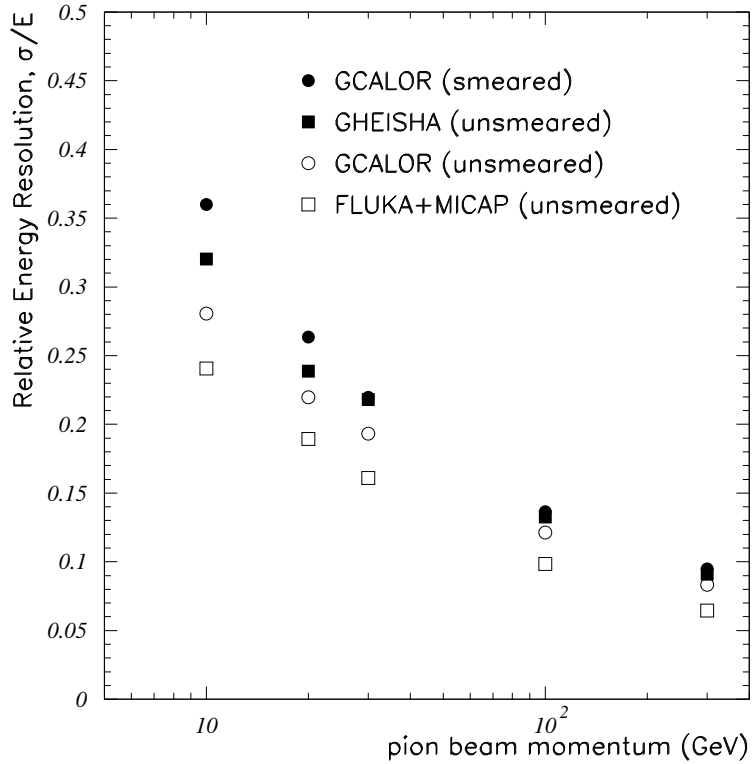}}
\caption{
Comparison of GEANT simulation for the pion energy resolution
of the crystal ECAL+HCAL combined system
using various generators (GHEISHA, GCALOR, and GFLUKA-MICAP)
of hadron showers. The test beam data is represented best by the 
GCALOR simulation, which include all test beam detector effects
and is labeled as "smeared". The GCALOR simulation labeled
as "unsmeared" is expected to represent the experimental situation in the
CMS experiment.}
\label{sk-96hf-97-0008}
\end{center}
\end{figure}

  Comparisons of the fractional energy resolution simulated by  GCALOR 
  and other GEANT hadron simulators 
  are shown in Figure~\ref{sk-96hf-97-0008}.
  The simulations based on GHEISHA  
  predict  somewhat worse resolutions, while simulations based on
  GFLUKA+MICAP predict much better resolutions than
  predicted with GCALOR and  observed in the data.

\section{Conclusions}

  Comprehensive tests of the 
  performance of prototype CMS central hadron 
  calorimeters have been done in the
   H2 and H4 beamlines at CERN.
   Data were taken with both a stand-alone HCAL calorimeter
   and with HCAL in combination with an upstream ECAL calorimeter.
 
  One of the primary objectives of the HCAL test beam studies 
  was to  investigate the calorimeter performance in the presence
  of perpendicular and parallel magnetic fields. 
  A high magnetic
  field changes the response of the calorimeter in two different ways: 
  (1) the  overall light yield of the scintillator
  is increased in a high magnetic field and (2) the field
  affects the observed energy deposition of electromagnetic 
  components of showers when it is parallel to the scintillator
  plates (i.e. perpendicular to the particle direction).
  For a collider experiment with a solenoid magnet, the magnetic field
  is parallel to the calorimeter plates in the central part of the detector
  (barrel configuration) and is perpendicular to calorimeter
  plates in the large $\eta$ region ( endcap configuration).

When the magnetic field
is perpendicular to the scintillator
planes (endcap configuration), only an intrinsic increase 
of the light yield of the scintillator of approximately 5-8\%
is observed relative to the case with
no magnetic field. This effect leads to the same
overall increased response of the calorimeter to muons, electrons, pions,
and $\gamma$ rays from a radioactive source. Therefore a calibration source
can be used to track and correct for this effect.

An additional geometric effect leads
to an increased response of the calorimeter to
the electromagnetic component of showers. This effect occurs
in the case in which the magnetic field lines are parallel to the
scintillator planes (barrel configuration)
and originates from an increase in the geometrical path length
of low energy electrons in a magnetic field. 
Since this effect is not present for the calibration source,
$\it{in~situ}$  (B field on)  calibration is required for hadron barrel
calorimeter.

The size of this effect is 
approximately
proportional to the strength of the B field and
depends on the detailed structure and composition
of the absorber and scintillator planes. 
However,
the effect  is expected to be present 
in any sampling calorimeter (independent of readout technology)
situated in a magnetic field which is parallel to 
the readout planes (barrel configuration).

Due to the non-compensating nature of the 
 lead tungstate crystal ECAL
the  linearity and energy resolution
of the  combined crystal ECAL+HCAL is worse than for HCAL alone.
Therefore, improvements in the linearity and resolution using
 weighting of the two longitudinal readouts (H1 and H2) of HCAL
 have been investigated.
Using a passive weighting, the fractional energy resolution of 
the combined crystal ECAL+HCAL calorimetric system 
is described by the
function $\sigma_{E}/E$=122\%/$\sqrt{E}$ $\oplus$ 5\%.
This performance is achieved by
a novel longitudinal segmentation of HCAL with
H1 being a single layer immediately behind ECAL. 
We conclude that for  combined ECAL+HCAL calorimeter systems with an
ECAL with a large e/h, a very thin first depth segment in
HCAL can be  used to largely correct for the
resultant non-linearity and degradation in energy resolution.
Monte Carlo studies of the CMS detector
in a collider enviornment  indicate that 
with the above performance, the
energy resolution for jets
is not dominated by the  energy measurement in the hadron calorimeter,
but by other fluctuations which are inherent in  jets~\cite{lhcc_ref}.

We find that the average longitudinal hadron shower profiles extend past the
inner HCAL located inside the
magnetic coil. 
Therefore,  the CMS central calorimeter design includes instrumenting
an outer calorimeter (HO)  outside the coil to measure the energy
in the tail of high energy hadronic showers.
The addition of HO leads to a total ECAL+HCAL+HO depth of
at least 9.5 $\lambda_{INT}$ for almost the entire $\eta$ range
spanned by the CMS hadron calorimeters.

Various GEANT based simulations  predict the impact of
the performance of HCAL on  a variety of potential physics searches
in the CMS detector. These simulations
extrapolate the performance of HCAL from the 
test beam configuration to the final configuration
chosen for CMS HCAL design. 
The Monte Carlo programs have been shown to successfully simulate various
test beam setups, including hadron shower
calorimetry in a setting with both 
crystal and copper-scintillator detectors in a strong
magnetic field. 
The best description of the
data is provided by GEANT with the GCALOR module for the generation of
hadron showers.

\section{Acknowledgments}

We thank the CERN staff and the staff of
the participating institutions for their
vital contributions to this experiment. We particularly  
acknowledge efforts of Maurice Haguenauer and Jean Bourotte (H4 beamline),
Pascal Petiot, and Gerhard Waurick (H2 beamline), Brian Powell (3T magnet),
Albert Ito (PMTs), as well as Anatoly Zarubin (DAQ).

\clearpage
\begin{table}
\begin{center}
\vspace{0.5cm}
\begin{tabular}{l|c|c} \hline\hline
& &  \\
Layer number& Absorber thickness & Scintillator thickness \\ \hline
& & \\
ECAL 1-10     & 1.6 cm Pb &  6 mm SCSN-38 \\
HCAL 1-9      & 5 cm Cu   &  4 mm SCSN-81 \\
HCAL 10-20    & 10 cm Cu  &  4 mm SCSN-81 \\
& & \\
\\ \hline\hline
\end{tabular}
\caption{Longitudinal segmentation of ECAL and HCAL
in the H2(1995) setup. Here a Pb-scintillator sampling calorimeter
is used for the ECAL section.}
\label{h2_1995}
\end{center}
\end{table}

\clearpage

\begin{table}
\begin{center}
\vspace{0.5cm}
\begin{tabular}{l|c|c} \hline\hline
& &  \\
Layer number& Absorber thickness & Scintillator thickness \\ \hline
& & \\
HCAL 1     & 2 cm Cu &  4 mm SCSN-81 \\
HCAL 2-10  & 3 cm Cu &  4 mm SCSN-81 \\
HCAL 11-19 & 6 cm Cu &  4 mm SCSN-81 \\
HCAL 20    & 8 cm Cu + 29 cm Al &  4 mm SCSN-81 \\
HCAL 21    & 8 cm Cu &  4 mm SCSN-81 \\
HCAL 22    & 8 cm Cu &  4 mm SCSN-81 \\
HCAL 23    & 10 cm Cu &  4 mm SCSN-81 \\
HCAL 24    & 10 cm Cu &  4 mm SCSN-81 \\
& & \\
\\ \hline\hline
\end{tabular}
\caption{Longitudinal segmentation of  HCAL
in the H4(1995) setup. Here the ECAL module consists of a 7$\times$7 matrix
of PbWO$_4$ crystals.}
\label{h4_1995}
\end{center}
\end{table}

\clearpage

\begin{table}
\begin{center}
\vspace{0.5cm}
\begin{tabular}{l|c|c} \hline\hline
& &  \\
Layer number& Absorber thickness & Scintillator thickness \\ \hline
& & \\
HCAL 1     & 2 cm Cu &  4 mm SCSN-81 \\
HCAL 2-7   & 3 cm Cu &  4 mm SCSN-81 \\
HCAL 8-21  & 6 cm Cu &  4 mm SCSN-81 \\
HCAL 22-27 & 8 cm Cu &  4 mm SCSN-81 \\
& & \\
\\ \hline\hline
\end{tabular}
\caption{Longitudinal segmentation of  HCAL
in the H2(1996) setup. Here, the ECAL module consists of a 7$\times$7 matrix
of PbWO$_4$ crystals. }
\label{h2_1996}
\end{center}
\end{table}

\clearpage
\clearpage

\clearpage

\clearpage

\clearpage

\clearpage

\clearpage

\clearpage
\clearpage
\clearpage

%
%
%

\clearpage

\clearpage

\clearpage

\clearpage

\clearpage

\clearpage

\clearpage

\clearpage

\clearpage

\clearpage

\clearpage

\clearpage

\clearpage

\clearpage

\clearpage

\clearpage

\clearpage

\clearpage

\clearpage

\clearpage

\clearpage

\clearpage


\clearpage

\clearpage

\clearpage

\clearpage

\clearpage

\clearpage
%
%
\clearpage

\clearpage
%

\clearpage

\clearpage

\clearpage

\clearpage

\begin{thebibliography}{999}
\bibitem{alloy} The absorber plates are composed of copper alloy (brass)
with the following composition: 59\% Cu, 39\% Zn, 1\% Fe, 1\% Mn.
\bibitem{tyvek} Tyvek B1060 (registered trademark of DuPont Co.) is
 a thin sheet of high density white polyethylene.
\bibitem{tedlar} Tedlar TCC15BL3 (registered trademark of DuPont Co.)
is a black opaque film of polyvinyl fluoride polymer.
\bibitem{nutev_lucy} The correction for
the  longitudinal energy leakage for a finite length calorimeter is 
estimated using CCFR and NuTeV test beam measurements of 
hadron showers as a function of calorimeter depth.
W. Sakumoto and L. de Barbaro, private communications.
\end{thebibliography}
\end{document}